\documentclass[fleqn,usenatbib]{mnras}

\usepackage{newtxtext,newtxmath}

\usepackage[T1]{fontenc}

\DeclareRobustCommand{\VAN}[3]{#2}
\let\VANthebibliography\thebibliography
\def\thebibliography{\DeclareRobustCommand{\VAN}[3]{##3}\VANthebibliography}

\usepackage{graphicx}	
\usepackage{amsmath}	

\usepackage{xcolor}
\usepackage[normalem]{ulem} 

\definecolor{arpcolor}{RGB}{0, 204, 204	}

\definecolor{edccolor}{RGB}{93, 63, 211	}

\title[Milky Way structure using live potentials]{Self-consistent modelling of the Milky Way structure using live potentials}

\author[E. Durán-Camacho et al.]{
Eva Durán-Camacho,$^{1}$\thanks{E-mail: durancamachoe@cardiff.ac.uk (EDC)}
Ana Duarte-Cabral,$^{1}$
Alex R. Pettitt,$^{2}$
Robin G. Tre{\ss},$^{3}$
Paul C. Clark,$^{1}$
\newauthor{}
Ralf S. Klessen,$^{4,5}$ 
Kamran R. J. Bogue,$^{6}$
Rowan J. Smith,$^{6,7}$
and Mattia C. Sormani$^{4}$
\\
$^{1}$School of Physics and Astronomy, Cardiff University, Cardiff, CF24 3AA, UK\\
$^{2}$Department of Physics and Astronomy, California State University, Sacramento, 6000 J Street, Sacramento, CA 95819-6041, USA\\
$^{3}$Institute of Physics, Laboratory for galaxy evolution and spectral modelling, EPFL, Observatoire de Sauverny, Chemin Pegais 51, 1290 Versoix, Switzerland\\
$^{4}$Universit{\"a}t Heidelberg, Zentrum fur Astronomie, Institut f{\"u}r Theoretische Astrophysik, Albert-Ueberle-Str. 2, 69120 Heidelberg, Germany\\
$^{5}$Universit{\"a}t Heidelberg, Interdisziplin{\"a}res Zentrum fur Wissenschaftliches Rechnen, Im Neuenheimer Feld 205, 69120 Heidelberg, Germany\\
$^{6}$School of Physics and Astronomy, University of St Andrews, North Haugh, St Andrews, KY16 9SS UK\\
$^{7}$Jodrell Bank Centre for Astrophysics, Department of Physics and Astronomy, University of Manchester, Oxford Road, Manchester M13 9PL, UK}

\date{Accepted 2024 June 5. Received 2024 May 11; in original form 2023 May 19}

\pubyear{2024}

\begin{document}
\label{firstpage}
\pagerange{\pageref{firstpage}--\pageref{lastpage}}
\maketitle

\begin{abstract}

To advance our understanding of the evolution of the interstellar medium (ISM) of our Galaxy, numerical models of Milky Way (MW) type galaxies are widely used. However, most models only vaguely resemble the MW (e.g. in total mass), and often use imposed analytic potentials (which cannot evolve dynamically). This poses a problem in asserting their applicability for the interpretation of observations of our own Galaxy. The goal of this work is to identify a numerical model that is not only a MW-type galaxy, but one that can mimic some of the main observed structures of our Galaxy, using dynamically evolving potentials, so that it can be used as a base model to study the ISM cycle in a galaxy like our own. This paper introduces a suite of 15 MW-type galaxy models developed using the {\sc arepo} numerical code, that are compared to Galactic observations of $^{12}$CO and \ion{H}{I} emission via longitude-velocity plots, from where we extract and compare the skeletons of major galactic features and the terminal gas velocities. We found that our best-fitting model to the overall structure, also reproduces some of the more specific observed features of the MW, including a bar with a pattern speed of $30.0 \pm 0.2$ km\,s$^{-1}$\,kpc$^{-1}$, a bar half-length of $3.2 \pm 0.8$\,kpc. Our model shows large streaming motions around spiral arms, and strong radial motions well beyond the inner bar. This model highlights the complex motions of a dynamic MW-type galaxy and has the potential to offer valuable insight into how our Galaxy regulates the ISM and star formation. 

\end{abstract}




\section{Introduction}
\label{sec:intro}

Star formation (SF) has historically been assumed to be universal, mainly due to the existence of a number of observational characteristics linked to the star formation process that seem to be invariant. One such universal aspect is the Kennicutt-Schmidt law - the direct relation between the gas surface density and the surface density of the star formation rates (SFR) that holds for several orders of magnitude and many different types of galaxies \citep[e.g.][]{schmidt1959rate,kennicutt1989star}, suggesting that the rate of star formation depends only on the amount of (dense) gas available, regardless of the properties of that dense gas itself. However, when resolving individual regions within a galaxy, such relations become less tight, with large intrinsic scatter appearing \citep[e.g.][]{leroy2013molecular,pessa2022variations}. The question then becomes what drives the scatter within a galaxy - is it the chaotic nature of the interstellar medium (ISM), or is some of that scatter controlled by the larger-scale dynamics of the host galaxies?

While in nearby galaxies, resolving the SF process within clouds is still at the limit of observational capabilities, studies of SF in the Milky Way (MW) have been able to reveal the exquisite detail of star forming regions \citep[e.g.][]{arzoumanian2011characterizing}, whose properties can be linked to the SF history of the clouds \citep[][]{schneider2010dynamic,duarte2011cloud,luong2011w43,peretto2013global,longmore2013variations}. However, linking how those variations in cloud properties (and subsequent SF) relate to the potential effect of the larger-scale dynamics of the Galaxy is complicated by our inability to unequivocally associate clouds with their environment, due to the fact that we are located within the galactic disc and only can access an edge-on perspective.

Our understanding of the Milky Way's structure remains uncertain, largely due to the challenges in creating a 3D map of the Galaxy \citep{elmegreen1985spiral}. Nevertheless, there is consensus that the Milky Way is a spiral-armed disc \citep{oort1958galactic, kerr1961galactic, georgelin1976spiral, dame1986largest, kolpak2003resolving, xu2016local}, 
although the precise morphology of these arms remains debated \citep[as highlighted in e.g.][]{xu2018spiral}. 
Initial studies proposed a 2-arm structure \citep{weaver1970spiral,francis2012evidence}, but the limited evidence for stellar spiral arms, as indicated by data such as GLIMPSE \citep{benjamin2005}, has led to alternative models proposing a 2+2 spiral structure \citep[e.g.][]{vallee2008new,efremov2011spiral,reid19}. A leading theory to reconcile these models suggests a dual approach: a four-arm structure predominantly of gas, alongside a two-arm structure composed mainly of stars \citep{drimmel2000evidence,martos2004galactic,steiman2010cobe}.

Modern observational efforts have brought exciting new data that allow the means to test galactic models, among which the data from the \textit{Gaia} mission stands out \citep[e.g][]{prusti2016gaia}. \cite{katz2018gaia} provide a first look into the potential of the mission using data from the \textit{Gaia} Data Release 2 (DR2), allowing for the creation of 3D velocity and velocity dispersion maps with unprecedented accuracy and spatial resolution. \cite{xu2021local} expand the results for the spiral arms obtained from the Very Long Baseline Interferometry (VLBI) maser data using the \textit{Gaia} DR3. They find new OB-type stars and based on their clumped distribution they reach the conclusion that the Galaxy spiral structure is irregular. Therefore, the overall structure of the Galaxy remains uncertain as even more new features are being found with the help of these new data releases.

Recent data from the \textit{Gaia} mission have significantly advanced our understanding of Galactic structures \citep[e.g.][]{prusti2016gaia}. \textit{Gaia} Data Release 2 (DR2) provided detailed 3D velocity maps, significantly enhancing spatial resolution and accuracy \citep{katz2018gaia}. Further insights from \textit{Gaia} DR3 continued to refine our understanding of the Galaxy's asymmetrical disk and spiral arms, suggesting an irregular and transient nature of the Milky Way's spiral structure \citep{xu2021local, castro2021}, with complex dynamical patterns \citep{khali2023,kawata2023,drimmel2023}. These ongoing discoveries from each \textit{Gaia} data release contribute to an ever-evolving picture of the Galaxy, revealing new complexities and challenging existing models.

The evaluation of the kinematics of the gas has been realised via longitude-velocity (\textit{l-v}) maps using different tracers. The key tracers used to track the galactic morphology are \ion{H}{I} \citep[e.g.][]{kulkarni1982atomic,kalberla2009hi} and $^{12}$CO emission \citep[e.g.][]{dame1986largest,dame2001milky,sawada2001tokyo}, which tend to indicate the presence of four main spiral arms. In the advent of higher resolution surveys of the galactic plane in optically thinner molecular lines \citep[e.g.][]{schuller2021sedigism, rigby2016chimps,colombo2021oghres}, we begin to resolve and discern the gas emission with greater detail, but the spiral structure as seen in the gas, appears to be more flocculent in nature than proposed by earlier works \citep[e.g.][]{duarte2021sedigism,colombo2022sedigism}.

The existence of a Galactic bar towards the Galactic Centre is well-supported by numerous observations, initially deduced via gas velocities \citep{cohen1976observations} and subsequently confirmed through photometry \citep{blitz1991direct} and star counts \citep{weinberg1992detection}. \citet{peters1975} proposed an elliptical bar model based on the \ion{H}{I} velocities. Findings from  COBE \citep{weiland1994} provided compelling evidence for the photometric asymmetry generated by the bar, and was subsequently followed by studies like \cite{stanek1994}, who first demonstrated a distance asymmetry with star counts. The recent study from \citet{drimmel2023} uses the latest DR3 to reveal a wealth of kinematic information, with prospects to enhance our knowledge of non-axisymmetric structures in the Galaxy.

Earlier studies indicated varying bar lengths and orientations \citep[e.g.][]{binney1997photometric,bissantz2002spiral} suggesting a bar with a half-length of $\sim 3.5$\,kpc and orientation of $\theta=20^{\circ}$. In contrast, \cite{Lopez-Corredoira2007} proposed a longer bar of $\sim 3.9$\,kpc  half-length at $\theta \sim 43^{\circ}$, while \cite{hammersley2000detection} observed a bar with a radii extending  $\sim 4$\,kpc at $\theta \sim 43^{\circ}$ using near-infrared data. Subsequently,  \citet{cabrera2007tracing} argue for two distinct bar structures, a long thin bar contained within the plane and orientation of $\theta=43^{\circ}$, and a triaxial boxy bulge oriented $\theta=12.6^{\circ}$. \cite{wegg2015structure} described a two-component long bar with estimated orientation angles of $28-33^{\circ}$, emphasizing a "thin" and "super-thin" structure, and a half length of $4.6 - 5$\,kpc. They contrast this with a 3D inner bulge/bar component at about $\sim 27^{\circ}$ \citep{wegg2013mapping}.  More recent observational studies found values for the pattern speed between $37-41$\,km\,s$^{-1}$\,kpc$^{-1}$ and bar orientations between $20-28^{\circ}$ \citep[e.g.][]{Bovy2019,clarke2019milky,queiroz2021milky}. The study from \cite{drimmel2023} used data from the \textit{Gaia} DR3 and found a bar orientation of $\sim20^{\circ}$ and pattern speed of $\sim38$\,km\,s$^{-1}$\,kpc$^{-1}$.

On the numerical side, \cite{pettitt2014morphology} in their hydrodynamical simulations of the interstellar medium (ISM) using smoothed particle hydrodynamics (SPH) conduct a parameter search with analytic prescriptions for the galactic bar potential, and find that their best combined model (arms+bar) occur with a pattern speed ($\Omega_{p}$) of $\sim 50-60$ km\,s$^{-1}$\,kpc$^{-1}$ and a bar orientation of $\sim 45^{\circ}$. On the other hand, \cite{li2022} use high resolution simulations with a barred potential based on observations in agreement with the boxy/peanut bulge hypothesis. Their results provide a good fit to the observed bar, with a bar pattern speed that is in the range of $37.5-40$ km\,s$^{-1}$\,kpc$^{-1}$. However, numerical studies focusing on the inner Milky Way, rather than encompassing the entire galactic disc, suggest pattern speed values as low as $30-40$ km\,s$^{-1}$\,kpc$^{-1}$ \citep[e.g.][]{weiner1999properties,sormani2015recognizing,wegg2015structure, portail2017dynamical,clarke2019milky, clarke2022,sormani2022stellar}.

With the observational challenges of getting a clearer view of how the ISM in the Milky Way is affected by its galactic structures, we approach this problem numerically. In order to model the spiral/bar features in the gas of the Milky Way using numerical simulations, there are two possibilities regarding how the galactic (dark matter and stellar) potential is represented. The most common approach is to set up an analytical gravitational potential for the gas to follow \citep[e.g][]{dobbs2013exciting,smith2020cloud}, which typically involves including spiral and bar patterns that rotate with a fixed angular velocity, i.e. as a solid body. There have been many simulations that use different techniques, such as those based on Eulerian grids \citep[e.g.][]{weiner1999properties}, particle-based Lagrangian methods where external potentials have been imposed \citep[e.g.][]{lee1999smoothed,pettitt2014morphology}, or moving-mesh codes with external potentials \citep[e.g.][]{tress2020simulations,sormani2018theoretical,hatchfield2021dynamically}.

The other approach to model spiral galaxies is to use stellar and dark matter particles to recreate the galactic potential, and generate the galactic structure self-consistently through the dynamical evolution of the system (i.e. so-called ``live potentials''). Large-scale over-densities in the galactic disc, like the spiral arms, or bars, should appear naturally from the gravitational and hydrodynamical interaction of the different components of the galaxy. There have been a number of studies that use live potentials to model spiral galaxies \citep[e.g.][]{baba2017eventful,tress2020simulations,pettitt2015morphology,iles2022differences}, and those models show that the response of the gas to the spiral arms is different in live potentials compared to the imposed potentials, as the arms are more dynamic and usually transient, not restricted to solid-body like rotations with a fixed pattern speed with radius \citep[e.g.][]{dobbs2014dawes,sellwood2019spiral,pettitt2020young}. While analytical potentials have a clear benefit of being highly tailored and controlled, the forces generated by the rigid external potential are imposed on the gas, whereas a live disc allows self gravity to interact both ways and induce a more realistic response of the gas.

Regardless of the type of potential used, most of the numerical works related to the Milky Way, however, typically set out to produce a Milky Way-type galaxy, rather than effectively attempting to fit the observed structure of the Milky Way, which requires the creation and subsequent analysis of \textit{l-v} maps \citep[e.g.]{fux19993d,wada2011interplay,li2016gas}. One exception is the work of \citet{pettitt2014morphology,pettitt2015morphology} who produce \textit{l-v} maps from their models and quantitatively compare their results with observations via a $\chi^{2}$-like statistic. Of particular relevance to our work, are the models with live potentials from \cite{pettitt2015morphology}, which can reproduce the observed spiral arm features and the gas dynamics of the Galaxy, but fail to reproduce an inner bar.

In this study, we aim to reproduce the structure of the Milky Way using numerical models with live potentials, by extending the parameter space probed by \cite{pettitt2015morphology}. We explore the evolution of 15 different models with varying initial stellar masses and bulge fractions, in order to assess which configuration is able to better reproduce the observed Milky Way structure. We compare with observations via \textit{l-v} maps, utilising a number of different statistical metrics. We investigate more specific features of our best model such as the spiral arms pattern, the galactic bar and its pattern speed, and the dynamics of the spiral arms at different radii. We then briefly compare with observations, assessing how good our best model is at reproducing these structures. Our best model will then serve in the future as the starting setup for a high resolution simulation with chemistry, feedback and star formation. It will allow us to track the evolution of the ISM in detail, in a model that is not only MW-like, but that attempts to mimic the large-scale dynamics of the Milky Way in a self-consistent way, key to understanding the effect of spiral structures of our own Galaxy on the SF within.

This paper is organised as follows. In Section~\ref{sec: numerical models} we describe the different methods used: the numerical code and the setup of the initial conditions. In Section~\ref{Section: Observations}, we review the observations used to constrain our simulations. In Section~\ref{sec: galactic features} we explain the techniques required to extract the different features of the model Galaxy, and in Section~\ref{Section: Best Fit} we present the results from the different statistical tests. In Section~\ref{Section: Properties} we examine the main galactic features of our best simulation compared to observations, and finally, we summarise our results in Section~\ref{Section: Conclusions}.

\section{Simulation methodology}
\label{sec: numerical models}

\subsection{Numerical method}
\label{sec:overview}

For our simulations, we use the moving-mesh hydrodynamical code {\sc arepo} \citep{springel2010pur} to simulate Milky Way-type spiral galaxies. Here, we give only a brief description of the main features of the code -- a more detailed explanation can be found in \cite{springel2010pur}, \cite{pakmor2016improving} and the public code repository\footnote{{\sc arepo} and relevant documentation can be found at \url{https://arepo-code.org/}.}. 

{\sc arepo} combines the accuracy of grid-based methods with the adaptivity and Galilean invariance of smoothed particle hydrodynamic (SPH) techniques, using a mesh defined by the Voronoi tessellation of a set of discrete points, which move with the velocity of the local flow.  For the gravity treatment, {\sc arepo} benefits from an improved tree-based approach from the {\sc GADGET-2} code \cite{2005MNRAS.364.1105S}.  It has successfully been used in the past on a variety of scales, being first designed for cosmological simulations at different epochs \citep[e.g.][]{marinacci2015large,weinberger2016simulating,pillepich2018simulating,pillepich2018first,springel2018first,pillepich2018first, nelson2019first}. {\sc arepo} has also been used to reproduce galaxies within a cosmological environment via the `zoom-in' technique \citep[e.g.][]{grand2016vertical,gomez2017lessons,simpson2018quenching,liao2019ultra,grand2019gas,van2021effect,garrison2018origin}, as well as individual isolated galaxies \citep[e.g.][]{smith2014co,glover2016co,sormani2018theoretical,sormani2019geometry,tress2020simulations,reissl2020synthetic}, and down to star formation inside individual molecular clouds and filamentary structures \citep[e.g.][]{clark2015does,smith2014nature,bertram2016synthetic,clark2019tracing,smith2020cloud,clarke2020hierarchical,whitworth2022molecular}.

As the main interest of this study is the dynamical evolution of the stars in order to form spiral arms and bars, we do not include the gas self-gravity in our calculations. This means that while the gas only feels the gravitational potential from the stars and dark matter, the stars and dark matter halo particles feel the entire potential, including the gas contribution. This has therefore no impact on the large-scale potential felt by the stars and reduces the overall computational time required, and the gas effectively effectively traces the stellar disc potential gradients. Gas is still required in our simulations for several reasons, such as facilitating an easier transitions to modelling with a full set of stellar/ISM physics or the well-defined features the gas traces when in contrast with stars or DM for a latter comparison with observations. Nevertheless, the total gas mass is only $\sim 20\%$ of the total stellar mass (and $\sim 1\%$ of the total mass of the galaxy), so the contribution of the gas to the overall galactic potential is minimal, and removing the gas self-gravity does not have a great impact on the global dynamics of the gas at kpc scales. Tests including self-gravity showed some gas accumulating in artificial clumps, but the overall structure was consistent with the current simulations.

The models presented here are three dimensional and the gas cells move freely with no imposed analytical potential. Instead, the galactic potential is recreated by the inclusion of stellar and dark matter particles, which evolve dynamically and self-consistently. Depending on the stability of the initial conditions, this will allow the creation of spiral structures and bars, although bars can take significantly longer to manifest than arm features (on Gyr timescales). Therefore, in order to allow the $N$-body part of the model to evolve dynamically for the duration needed for bar formation, while maintaining a reasonable computational time, we keep a relatively coarse resolution for the gas cells and impose an isothermal equation of state to avoid collapse associated with ISM cooling processes. The temperature, $T$, is controlled by the sound speed, $c_{s}$ defined in Eq.\,\ref{eq cs} for the isothermal equation of state, where $k_{B}$ is Boltzmann's constant, and $m$ is the mass of a single hydrogen atom.
\begin{equation}
    \label{eq cs}
     c_{s}=\sqrt{\frac{k_{B}T}{m}}
\end{equation} 
We adopt $c_{s}=10$ km\,s$^{-1}$, which corresponds to a temperature of $T \sim 10^{4}$ K. A further discussion of the equations solved by {\sc arepo} can be found in \cite{tress2020simulations}.

The gas resolution of our simulations is defined by a refinement criteria based on mass and volume of the gas cells. For our models we use a gas target mass of $1000$\,M$_{\odot}$, and we impose a limit to the volume of each cell. The minimum cell volume corresponds to $27$\,pc$^{3}$ (equivalent to a cube of $3$\,pc in side) and the maximum volume to $1.25\times10^{8}$\,pc$^{3}$, equivalent to a cube of $500$\,pc in side. The mass resolution used for each type of particle as well as the respective softening length are presented in Table \ref{tab:resolution}. The gravitational softening of gas cells adheres to {\sc arepo}'s adaptive softening scheme and changes throughout the simulation in accordance with changes in volume of the cells. 

Our models are simulated within a box of $100$ kpc side, and we use periodic boundary conditions for the gas. However, our box is big enough to avoid any boundary effects in the simulated galaxy. The total number of gas cells found in each of our simulations are in the range of $23-26$ million for the most evolved models. 

\begin{table}
	\centering
	\caption{Mass resolution and softening length of the different particle types. Gas cells follow {\sc arepo}'s adaptive softening scheme, hence here are included the minimum and maximum softening values marked with $\star$.}
	\label{tab:resolution}
	\begin{tabular}{lcc} 
		\hline
		 & Mass resolution ($M_{\odot}$) & Softening (pc)\\
		\hline
		Dark matter & $9 \times 10^{5}$ & 300 \\
		Stars  & $8 \times 10^{3}$ & 50  \\
		Gas & $1 \times 10^{3}$ & $(1-4)^{\star}$ \\
		\hline
	\end{tabular}
\end{table}

\subsection{Initial conditions}
\label{sec:initial conditions}

\begin{figure}
	\includegraphics[width=\columnwidth]{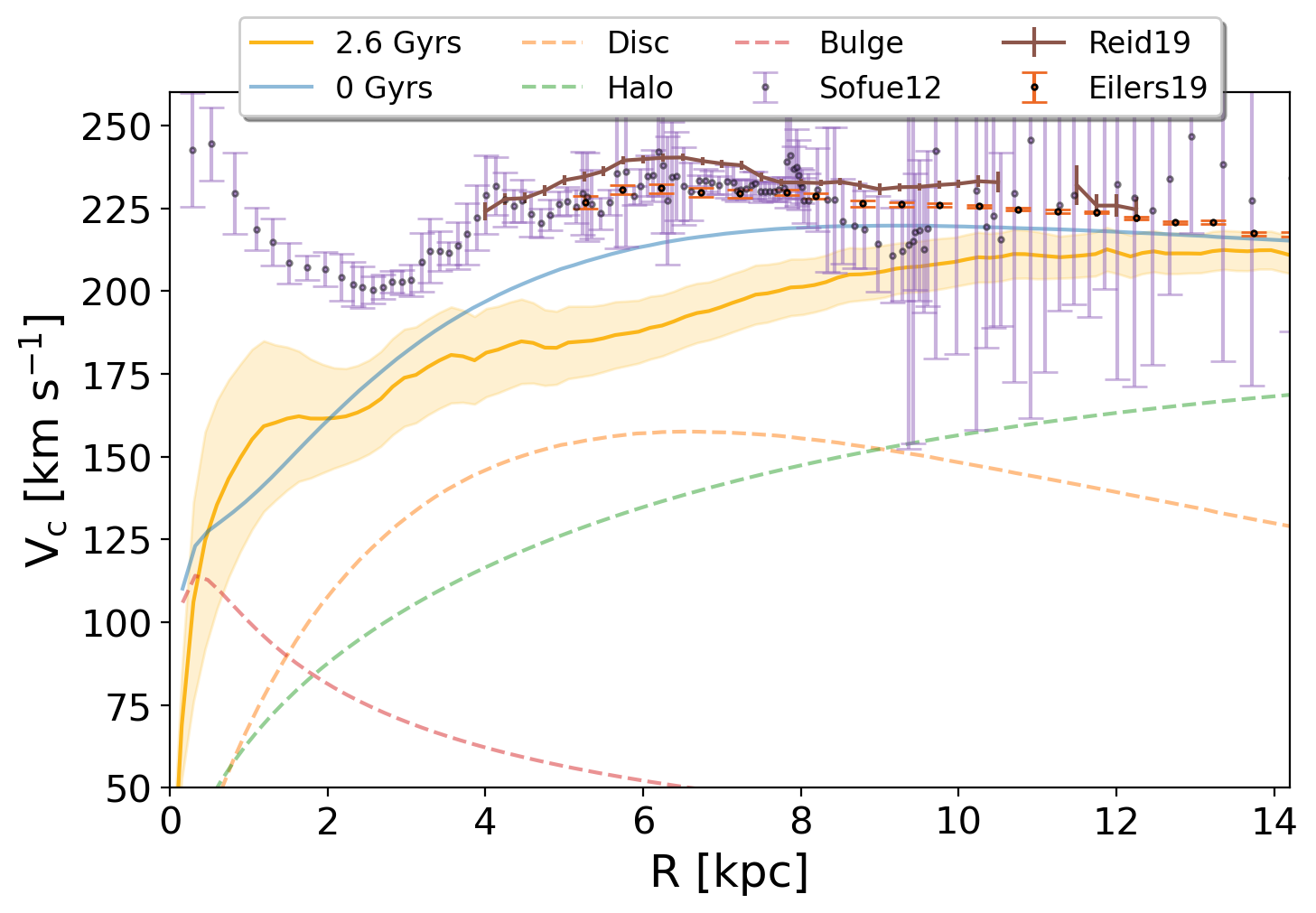}
    \caption{Example of the rotation curve for Model 4. The blue line indicates the initial rotation curve generated by the {\sc{makenewdisk}} code, with the contribution from the different galactic components shown in dashed lines: orange for the disc, green for the halo and red for the bulge. The solid yellow line represents the rotation curve of the same model at a later time, where the shaded area constitutes the 1$\sigma$ standard deviation (in velocity) of the data at each radial bin, at that later time. The observations from \protect\citet{sofue2012grand}, \protect\citet{eilers2019circular} and \protect\citet{reid19} all shown as black dots, but with light purple, brown and  orange error bars respectively.  Data for \protect\citet{sofue2012grand} has been corrected so that the circular velocity of the LSR matches that of \citet{eilers2019circular} at 229 km s$^{-1}$.}
    \label{fig:rot_curves_ICs}
\end{figure}

The initial conditions for our Milky Way models were set up using the moment-based code {\sc{makenewdisk}}, described in \cite{springel2005modelling}, and they consist of a spherically symmetric dark matter halo, a rotationally supported disc of gas and stars, and a central stellar bulge. Motivated by cosmological simulations, the dark matter halo has a mass distribution that peaks in the centre and drops at larger radii, following the \cite{hernquist1990analytical} profile, and whose analytical representation follows: 

\begin{equation}
\label{eq DM}
    \rho_{\mathrm{h}} (r) = \frac{M_{\mathrm{h}}}{2\pi}\frac{a}{r(r+a)^{3}}
\end{equation}

\noindent where, $M_{\mathrm{h}}$ refers to the total halo mass and $a$ is the scale length of the Hernquist profile for the halo. This distribution is in agreement with the well-known Navarro-Frank-White (NFW) profile \citep{navarro1996structure} in its inner parts, but has a faster decline outwards. The scale length is often given in terms of the halo concentration $c$, which allows a direct comparison between profiles, and it is defined as $c=r_{200}/r_{s}$, where $r_{200}$ is the radius at which the mean enclosed DM density is $200$ times the critical density over the scale length of the NFW profile. The relation between $a$ and $c$ is given by 

\begin{equation}
\label{eq c}
    a = r_{s} \sqrt{ 2[\ln(1+c)-c/(1+c)] }.
\end{equation}

A more in depth comparison of the NFW and Hernquist profiles can be found in the original reference. To set up each of our models, we explore a range of concentration indices, $c$, between $3-11$, and decide the final concentration from the global rotation curve generated, in comparison with the observational values from the works of \cite{sofue2012grand} and \cite{eilers2019circular}. 

Note that {\sc arepo} requires all space within the defined box to be populated with cells, i.e. no empty space is allowed as in Lagrangian codes. Therefore, some cells with very low density material are needed in order to facilitate the creation of the Voronoi tessellation. Hence, the total mass of the halo is in fact a combination of dark matter particles (with total mass $M_{\mathrm{h}}^{\mathrm{dm}}$) and gas (with mass $M_{\mathrm{h}}^{\mathrm{gas}}$).. For all models, we assume the gaseous halo mass to be $40\%$ of the gas mass in the disc, $M_{\mathrm{d}}^{\mathrm{gas}}$.  

The stellar bulge is also setup assuming spherical symmetry and a Hernquist profile with a scale length $b$, and total mass $M_{\mathrm{b}}$.

\begin{table*}
    \centering
    \begin{tabular}{ | *{12}{c|} } 
        \hline
         Model & $M_{\star}$ & $BF$ &  $\lambda$  &  $J_{d}$ &  $M_{d}^{\star}$ & N$_{d}^{\star}$  & $M_{b}^{\star}$ & N$_{b}^{\star}$ & $c$ & $M_{h}^{dm}$  & N$_{h}^{dm}$\\ 
         & [$10^{10}$ M$_{\odot}$] & [$\%$] &    &   & [$10^{10}$ M$_{\odot}$] & [$10^{6}$]  & [$10^{10}$ M$_{\odot}$] & [$10^{6}$] &  & [$10^{10}$ M$_{\odot}$]  &  [$10^{6}$]\\ 
        \hline
        \hline
        1  & 4.25 & 25 &  0.037  &  0.040 &  3.19 & 3.98 & 1.06 & 1.33 & 8  & 89.75  & 0.99\\ 
        2  & 4.25 & 20 &  0.037  &  0.042 &  3.40 & 4.25 & 0.85 & 1.06 & 9  & 89.75  & 0.99\\ 
        3  & 4.25 & 15 &  0.038  &  0.048 &  3.61 & 4.52 & 0.64 & 0.80 & 9  & 89.75  & 0.99\\ 
        4  & 4.25 & 10 &  0.039  &  0.049 &  3.83 & 4.78 & 0.43 & 0.53 & 9  & 89.75  & 0.99\\ 
        5  & 4.25 & 5  &  0.040  &  0.051 &  4.04 & 5.05 & 0.21 & 0.27 & 10 & 89.75  & 0.99\\ 
        \hline
        6  & 5.25 & 25 &  0.033  &  0.050 &  3.94 & 4.92 & 1.31 & 1.64 & 5  & 88.75  & 0.98\\
        7  & 5.25 & 20 &  0.033  &  0.053 &  4.20 & 5.25 & 1.05 & 1.31 & 5  & 88.75  & 0.98\\
        8  & 5.25 & 15 &  0.035  &  0.055 &  4.46 & 5.58 & 0.79 & 0.98 & 6  & 88.75  & 0.98\\
        9  & 5.25 & 10 &  0.035  &  0.059 &  4.73 & 5.90 & 0.53 & 0.66 & 6  & 88.75  & 0.98\\
        10 & 5.25 & 5  &  0.034  &  0.061 &  4.99 & 6.23 & 0.26 & 0.33 & 6  & 88.75  & 0.98\\
        \hline
        11 & 6.25 & 25 &  0.031  &  0.058 &  4.69 & 5.86 & 1.56 & 1.95 & 3  & 87.75  & 0.97\\
        12 & 6.25 & 20 &  0.031  &  0.062 &  5.00 & 6.25 & 1.25 & 1.56 & 3  & 87.75  & 0.97\\
        13 & 6.25 & 15 &  0.031  &  0.065 &  5.31 & 6.64 & 0.94 & 1.17 & 3  & 87.75  & 0.97\\
        14 & 6.25 & 10 &  0.033  &  0.068 &  5.63 & 7.03 & 0.63 & 0.78 & 4  & 87.75  & 0.97\\
        15 & 6.25 & 5  &  0.033  &  0.071 &  5.94 & 7.42 & 0.31 & 0.39 & 4  & 87.75  & 0.97\\
        \hline
        \hline
    \end{tabular}
    \caption{Parameters used to build the initial conditions for our different models with the {\sc{makenewdisk}} code: $M_{\star}$ is the total stellar mass and values are based of \protect\cite{pettitt2015morphology} best model and the mass profile from \protect\cite{cautun2020milky}; $BF$ is the stellar bulge mass fraction; $\lambda$ is the spin parameter \protect\citep[see][]{springel2005modelling};  $J_{d}$ is the disc spin fraction; $M_{d}^{\star}$ is the stellar mass in the disc; N$_{d}^{\star}$ is the number of stellar particles in the disc; $M_{b}^{\star}$ is the stellar mass in the bulge; N$_{b}^{\star}$ is the number of stellar particles in the bulge; $c$ is the halo concentration; $M_{h}^{dm}$ is the dark matter mass in the halo; and N$_{h}^{dm}$ is the number of dark matter particles in the halo. The number of particles are set by the imposed mass resolution and total masses.  The values for the disc scale height and bulge scale length are common to all values and are $0.29$ kpc and  $0.35$ kpc respectively.} 
    \label{table:1}
\end{table*}

The disc component (of both gas and stars), follows an exponential surface density ($\Sigma_{\mathrm{d}}$) profile as:

\begin{equation}
\label{eq exponential}
    \Sigma_{\mathrm{d}}(r)=\frac{M_{\mathrm{d}}}{2\pi h^{2}}\ {\rm exp}(-r/h)
\end{equation} 

\noindent where $h$ is the disc scale length, and $M_{\mathrm{d}}$ is the total mass in the disc, i.e. the sum of both gaseous ($M_{\mathrm{d}}^{\mathrm{gas}}$) and stellar ($M_{\mathrm{d}}^{\mathrm{\ast}}$) components\footnote{ The assumption that the gas follows the same exponential profile as the stellar disc is likely not appropriate for the MW. Nevertheless, we have tested the impact of redistributing the gas using a different surface density profile - such as a flat one - and found that the overall resulting structures (bar/spiral arms) were identical. Therefore, for the purpose of this paper, we have kept the original surface density profiles as per {\sc makenewdisc}. However, in future work - when including star formation and feedback - the gas surface density profiles will need to be adjusted in order to be a better representation of the MW's gas surface densities.}. In the z-direction it follows the same profile as in \citet{springel2005modelling} with a scale height given by $z_{0}$. 

The disc scale length is directly related to the spin parameter $\lambda$ and the fraction of angular momentum in the disc $J_{d}$, which assumes conservation of specific angular momentum of the material that forms the disc. Both parameters are inputs to the {\sc{makenewdisk}} code and are directly correlated with $M_{\mathrm{d}}$ and $h$ ( see \citealt{springel2005modelling} for details). The spin parameter $\lambda$ is  also dependent on the concentration of the dark matter halo, $c$. For our simulations, we adopt $h = 3$ kpc as per the work of \cite{pettitt2015morphology}. Hence, $\lambda$ is chosen so as to obtain the desired scale length for that specific $c$. 

 In terms of velocities, as explained in full detail in Section 2.3 of \citet{springel2005modelling}, the initial rotation of the disc is determined from the potential, and the vertical velocity dispersion of the stars in the disc is set by the specified disc scale height.   We adopt the default ratio as in \citet{springel2005modelling}. For our models, the resulting Toomre Q parameters of the discs are always above a value of $6$, thus they are largely Toomre stable throughout. For the bulge and halo, the velocities are setup such that the bulge will have no net rotation, and the halo has the same specific angular momentum as the disc.

A detailed description of all the parameters can be found in \cite{springel2005modelling}. 
Most of the parameters we use for setting up our Milky Way models were based on the work by \cite{pettitt2015morphology}, given that they use hydrodynamical + $N$-body simulations to represent the disc and bulge of the Milky Way with live potentials as well. We therefore take their best model (i.e. Bc/Normal model) as our starting point and explore a few variations on the stellar mass distribution, based on observational constraints of the total stellar mass of the Milky Way.

For all models, we assume a total gas mass in the disc of $M_{\mathrm{d}}^{\mathrm{gas}} = 8.6 \times 10^{9}$\,M$_{\odot}$, so the exponential surface density integrated within $13$\,kpc reaches $\sim 8 \times 10^{9}$\,M$_{\odot}$ as in \cite{pettitt2015morphology}. Their Bc/Normal model has a total stellar mass (disc $+$ bulge) of $M_{\mathrm{stars}} = 4.25 \times 10^{10}$ M$_{\odot}$, which is on the lower end of more recent observationally-derived values. For instance, the work by \cite{cautun2020milky} infers the Milky Way stellar mass profile by fitting observational data from the \textit{Gaia} DR2, and due to the large uncertainties in the observations, they find their best fit with a total stellar mass of $M_{\mathrm{stars}} \sim 5 - 6 \times 10^{10}$ M$_{\odot}$, depending on the assumed halo profile. Therefore, in this work we explore three initial stellar masses, $4.25 \times 10^{10}$\,M$_{\odot}$; $5.25 \times 10^{10}$\,M$_{\odot}$ and $6.25 \times 10^{10}$\,M$_{\odot}$, covering the range of observed values for the Milky Way. These values are also in agreement with the mass models found by \cite{mcmillan2016mass}.

In addition, we vary the stellar bulge fraction ($BF = M_{b}^{\star} / M_{\star}$) to explore the effects on the overall galactic structure. We start with $25\%$ similar to \cite{pettitt2015morphology}, and decrease in steps of $5\%$ until we reach the extreme $5\%$, where there is almost no bulge. Altogether, we generate a range of 15 different models that extend over a wide range of parameters to explore the combination that best reproduces the observations of our Galaxy. We enumerate our models from $1$ to $15$ with increasing initial stellar mass, and from larger to smaller bulge fraction (see Table.~\ref{table:1}).

Finally, the total mass of the galaxy (i.e. dark matter, gas and stars) is constant throughout all models and determined using
\begin{equation}
    \label{eq-Mt}
    M_{\mathrm{total}}=\frac{V_{200}^{3}}{(10GH_{0})},
\end{equation}

\noindent where $G$ is Newton's gravitational constant and $H_{0}$ Hubble's constant, and the virial velocity $V_{200} = 160$ $\mathrm{km}\ \mathrm{s}^{-1}$ is taken from \cite{springel2005modelling}. 

Figure~\ref{fig:rot_curves_ICs} shows an example rotation curve for Model 4.
 For context, we compare this model with observed values from \cite{eilers2019circular}, determined from 6D data from giant stars and for which the circular velocity at the Sun's location in the Local Standard of Rest (LSR) is assumed to be V$_{0} = 229\ \mathrm{km}\ s^{-1}$.  Additionally, we include the rotation curve form \citet{reid19}, derived from maser observations and independent of the LSR velocity assumption.  We have also included the data from \citet{sofue2012grand}, adjusting their initial assumption of V$_{0} = 200\ \mathrm{km}\ s^{-1}$ to align with the higher circular velocities of the LSR reported by \citet{eilers2019circular}. This adjustment uses a distance of $8.2$ kpc for the LSR, based on the results from \citet{abuter2019geometric}.  The initial rotation curve of Model 4, depicted in blue in Fig.~\ref{fig:rot_curves_ICs}, is generated using the {\sc{makenewdisk}} code. The dashed lines represent the contribution from the different galactic components: the disk in orange, the halo in green, and the bulge in red. The solid yellow line illustrates the model's rotation curve computed at a later time $t = 2.6$\,Gyr. The shaded area constitutes the 1$\sigma$ standard deviation (in velocity) of the data at each radial bin, for that later time.  Note that the effect of changing the bulge fraction affects the inner part of the galaxy, whereas the disc dominates towards larger radii. Therefore, for a given total stellar mass, the larger the $BF$, the stronger the peak in the rotation curve at $R<1$\,kpc. On the other hand, increasing the initial stellar mass reduces the halo mass, given that the disc gas mass and total mass of the galaxy are fixed parameters. Hence, for a constant $BF$, the larger the stellar mass, the faster the rotation curve decays at larger distances from the galactic centre, where it is dominated by the halo contribution. When compared to the observations (see in Figure.~\ref{fig:rot_curves_ICs} for reference),  our models tend to underestimate the velocities, especially at radii below 8\,kpc. We note, however, while the initial rotation curve is only slightly underestimated at those intermediate radii (by up to $\sim$10\%), the difference between the rotation curves of our models and the observations becomes more severe at later times, after the disc has settled - likely due to radial migration of mass, and perhaps linked to the lack of gas self-gravity (which promotes an initial redistribution of the gas outwards). Another notable difference is towards the galactic centre (i.e. $R<2$\,kpc ) in the initial conditions, but this aids to destabilise the central region and create the right conditions to generate a bar, which tends to be hampered by the presence of strong spherically symmetric and non-rotational components such as bulges and centrally concentrated halos \citep[see e.g.][and references in the introduction]{kataria2018study}.   With the evolution of the galaxy and the formation of a central bar, the rotation curve changes and we can see that the central region does reach higher velocities than in the original curve. However, as we will see in Sect.~\ref{Section:CMZ}, even with the bars, we still underestimate the velocities in the most central region ($R < 1$\,kpc), in which the motions of the MW are mostly dominated by $x_{2}$ orbits \citep[see e.g.][]{binney1991understanding,hatchfield2021dynamically}, which we do not resolve properly in all of our models.  We note, however, that the observed velocities inside a radius of $\sim 4$ kpc are unlikely to be correct due to the assumption of circular orbits in the bar region.

Summarising, we have $15$ different models and vary their initial stellar mass and bulge fractions to account for a diversity of possible large scale configurations. Each model has a fixed halo concentration, and spin parameter tailored to provide a disc scale length of $h = 3$ kpc. Disc scale height ($z_{0}$) and bulge scale length ($b$) are calculated in terms of $h$, and have final values for all models of $0.29$ kpc and $0.35$ kpc respectively. The total mass of the galaxy is constant throughout the sample, $M_{\mathrm{total}} = 9.52 \times 10^{11}$ M$_{\odot}$, as well as the disc gas mass, $M_{\mathrm{d}}^{\mathrm{gas}} = 8.6 \times 10^{9}$ M$_{\odot}$. We assume that the gas mass in the halo corresponds to $40 \%$ of the gaseous component of the disc, and thus it is a fixed quantity, $M_{\mathrm{h}}^{\mathrm{gas}} = 3.44 \times 10^{9}$ M$_{\odot}$. Table \ref{table:1} collects all the other different parameters for our range of models that are not predetermined. Example top-view plots of our different models can be seen in Appendix.~\ref{appendixB}.

\section{Observations}
\label{Section: Observations}

\begin{figure}
	\includegraphics[width=\columnwidth]{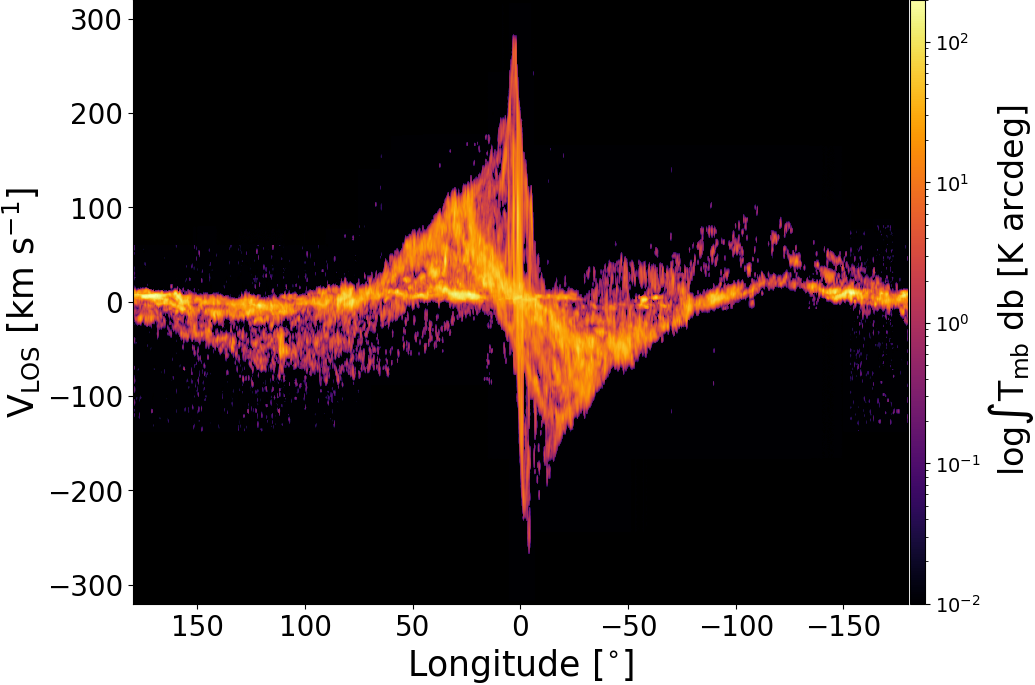}
    \caption{All-sky longitude-velocity map of the Milky Way extracted from the work by \protect\citet{dame2001milky}. This map was smoothed to a resolution of $\Delta v = 2$ km\,s$^{-1}$ in velocity and $\Delta l = 12'$ in longitude. The colourbar refers to the logged integrated brightness temperature of the  CO ($J=0-1$) transition.}
    \label{fig:Dame}
\end{figure}

\begin{figure}
	\includegraphics[width=\columnwidth]{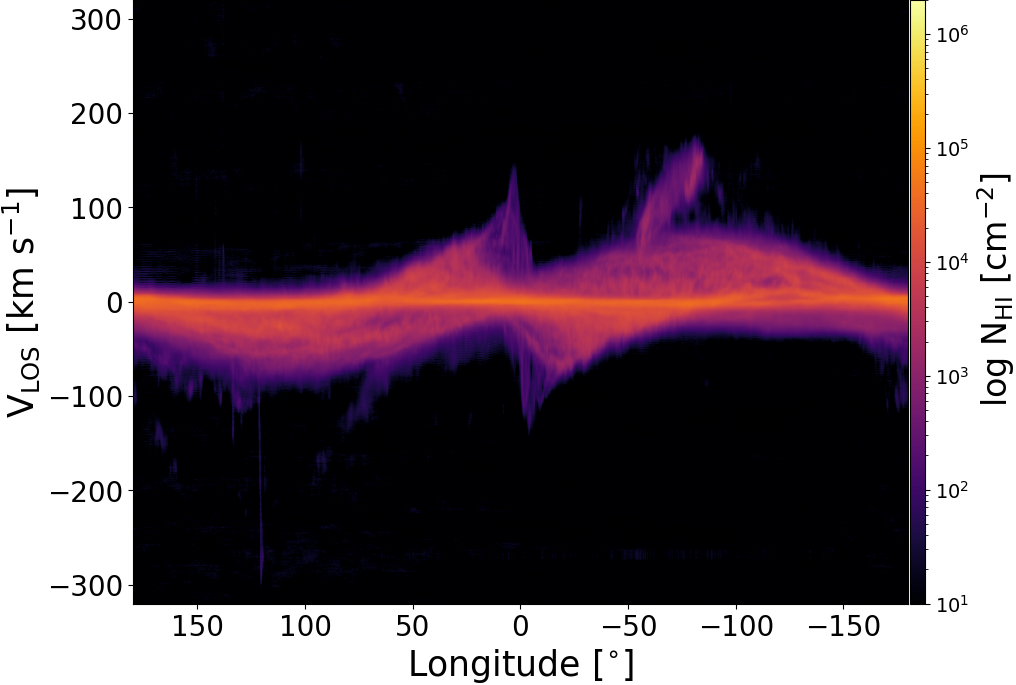}
    \caption{All-sky longitude-velocity map of \ion{H}{I} column density of the Milky Way. The data was extracted from the HI-4PI Survey \protect\citep{bekhti2016hi4pi} and plotted in a logarithm scale.}
    \label{fig:HI4PI}
\end{figure}

We make use of the $^{12}$CO emission cubes from \cite{dame2001milky}\footnote{Data retrieved from the Radio Telescope Data Center: \url{https://lweb.cfa.harvard.edu/rtdc/CO/CompositeSurveys/}} as the primary dataset of comparison with our models. We use them to obtain the main galactic features as seen in longitude-velocity space, as this is the gas tracer that sees the galactic spiral features with highest contrast. The \cite{dame2001milky} CO survey covers the entirety of the galactic plane, and the measurements were obtained with the CfA $1.2$ m Millimeter-Wave Telescope at the Harvard-Smithsonian Center for Astrophysics and its counter-part instrument at CTIO in Chile. This survey is formed by $488,000$ spectra that cover the Milky Way over a strip wide in latitude of  $4^{\circ}$ to $10^{\circ}$. For our study, we use the \textit{l-v} map that can be seen in Fig.~\ref{fig:Dame}, which was smoothed to a resolution of $\Delta v = 2$ km\,s$^{-1}$ in velocity and $\Delta l = 12'$ in longitude. 

In addition to the $^{12}$CO data, we also make use of \ion{H}{I} data to determine the terminal velocities of the Milky Way in \textit{l-v} space (see Section.~\ref{sec: terminal velocity}), as it traces the more diffuse gas of the outer Galaxy better than CO. For the \ion{H}{I} emission we use the HI-4PI Survey \citep[][]{bekhti2016hi4pi}, which contains data for $N_{H_{I}}$ column density as well as latitude, longitude and brightness temperature. The HI-4PI Survey is a combination of the first coverage of the Effelsberg–Bonn \ion{H}{I} Survey (EBHIS) and the third version of the Galactic All-Sky Survey (GASS). It has an angular resolution of $\vartheta_{fwhm} = 16.2$, and covers the complete Milky Way in longitude. All the data is publicly available\footnote{HI-4PI Survey: \url{http://cade.irap.omp.eu/dokuwiki/doku.php?id=hi4pi}}.

\section{Extraction of the main Galactic features}
\label{sec: galactic features}

In this section, we present the methods that we use to quantify how well our numerical models reproduce the observed galactic structures. In particular, we explain how we place ourselves inside the models in order to mimic the observer's perspective of the Milky Way (i.e. through \textit{l-v} plots), the techniques we use to identify the key galactic features, 
and the metrics used to quantify the how the models compare with observations.

\subsection{Construction of \textit{l-v} plots}
\label{sec:lv-plots}

The spiral patterns of galaxies are generated by the dynamical evolution of the different galactic components, and the larger stellar potential well inside a spiral arm aids the gas to concentrate, favouring the formation of larger complexes of molecular gas (giant molecular clouds). As such, molecular gas is typically a good tracer of the spiral pattern of galaxies \citep[e.g.][]{wright2001molecular, combes2007molecular, young1991molecular}, as it presents a higher contrast between arm and inter-arms regions than if observing the atomic gas (e.g. \citealt{querejeta2021stellar} where they used \ion{H}{I}). This spiral pattern can be seen as clear lines (and loops) in \textit{l-v} maps of gas tracers, such as those in \cite{dame2001milky}. For our models, given that we are simply interested in the position of the spiral arm tracks in \textit{l-v} space, and that we do not include chemistry or cooling in the models, we instead simply generate synthetic \textit{l-v} maps from the total gas mass along each line of sight.

The framework for the construction of \textit{l-v} plots from a 3D galaxy is explained in detail in \cite{binney2011galactic} and \cite{Binney1998}, as well as in \cite{pettitt2014morphology}. However, here we will give a brief summary of the process, as we adopt a more graphical approach.

\begin{figure}
	\includegraphics[width=\columnwidth]{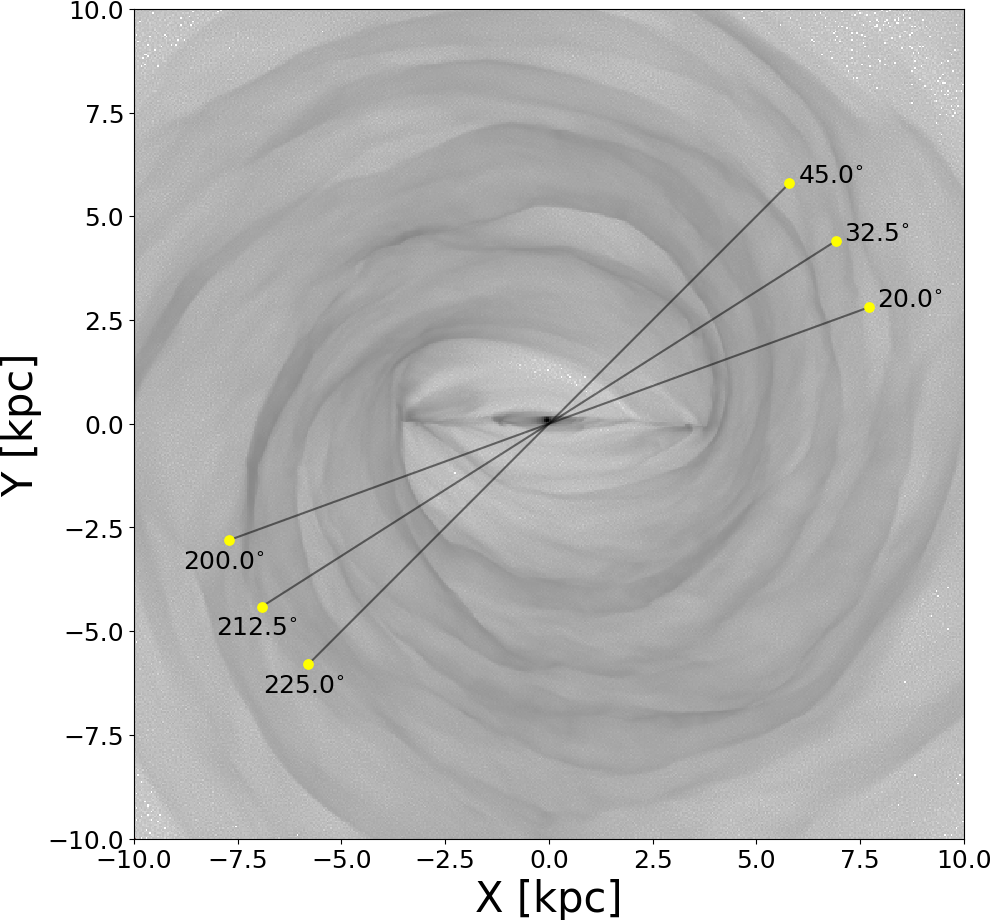}
    \caption{ Example of one of the top-down views of  Model $4$ at a time of $\sim 2.6$\,Gyr rotated so the bar is along the $x$-axis, and illustrating the range of angular positions in the box, $\phi_{obs}$, assumed for the Sun's position when creating the \textit{l-v} maps.}
    \label{fig:top_view}
\end{figure}

The first step in constructing an \textit{l-v} map is to adopt a position for the observer (i.e. the Sun's position). These coordinates can be defined by the distance from the galactic centre, or $R_{\mathrm{obs}}$, the angular position in the box within the disc with respect to the Sun-Galactic centre line $\phi_{obs}$, and the tangential velocity at the Solar position $V_{\mathrm{obs}}$. { The position of the galactic centre is determined for each time, as the centre of mass of the entire system (gas, stars and dark matter). For $R_{\mathrm{obs}}$}, we adopt an approximate value of $R_{\mathrm{obs}}\sim 8.2$ kpc, based on the study by \cite{abuter2019geometric}.  The azimuthal position of the observer is then essentially a free parameter. We first perform a Weighted Principal Component Analysis (WPCA, \citeauthor{bailey12} \citeyear{bailey12}) on the top-view column density of each model in our sample to determine the orientation of the Galactic bar. Afterwards, we rotate the model to align the bar along the $x$-axis and examine six different angles. Half of these angles are selected to ensure that the bar's orientation falls within the observed range of $20^{\circ}$ to $45^{\circ}$. Therefore we select angles spaced by $12.5^{\circ}$, and the remaining three angles are their mirrored counterparts, effectively adding $180^{\circ}$ to each of the initial three angles. We then produce an \textit{l-v} map for each observer position. An illustration of the different angles used can be seen in Fig. ~\ref{fig:top_view}. 

Using the galactic centre as the origin, the Sun position can be calculated in polar coordinates via $R_{\mathrm{obs}}$ and $\phi_{obs}$. Given the gas position and velocity from {\sc arepo} in Cartesian coordinates,  we then rotate the galaxy clockwise with respect to the galactic centre, such that the Sun is positioned at $y$=0 and $x$=$R_{\mathrm{obs}}$. We then obtain the longitude $l_{\mathrm{gas}}$ of each gas cell using Eq.~\ref{eq: longitude}:

\begin{equation}
    \label{eq: longitude}
    l_{\mathrm{gas}} =\frac{3 \pi}{2} - \mathrm{atan2}(y,x),
\end{equation}

\noindent where $x$ and $y$ are the gas coordinates in the rotated frame  with origin at the Sun's position.  We then obtain the velocity of the gas at the observer's position, $V_{\mathrm{obs}}$, as the mean velocity of all particles that fall within a range of radius of $R_{\mathrm{obs}}\pm 0.1$ kpc of the observer. We assume that the observer's motion is purely circular (i.e. we neglect any radial motions), and therefore we assume that its velocity has only a $y$-component in this frame. Finally, we project the cell velocity in the line-of-sight direction via a dot product of the cell velocity vector and the unity vector in the line of sight direction of the observer, obtaining $v_{\mathrm{los}}$. 

\subsection{Skeleton extraction}
\label{sec:skeletons}

We aim to trace the position of the spiral arms in the gas and carry out a direct comparison with observations, where the denser ridges of the spiral arms are best traced by the molecular gas, and in particular we select CO emission.  A more in depth comparison  of numerical and observational data, would ideally need a post-processing analysis of the numerical data with radiative transfer codes, in order to simulate the observed emission \citep[e.g.][]{duarte2015synthetic}. Nevertheless, given that the large-scale galactic features are mainly dominated by the large-scale dynamics of the stellar population rather than the local physics of the gas \citep[][e.g.]{pettitt2014morphology}, it is possible to do a feature comparison based on matching the fingerprints (or ``skeletons'') \citep{fux19993d,rodriguez-fernandex2008}, of our simulations to those in observations. As such, we extract the 'skeletons' of all \textit{l-v} maps { in order to highlight the position of the key structures as seen in the \textit{l-v} space in both the models, and in the observations.}

\begin{figure*}
	\includegraphics[width=\textwidth]{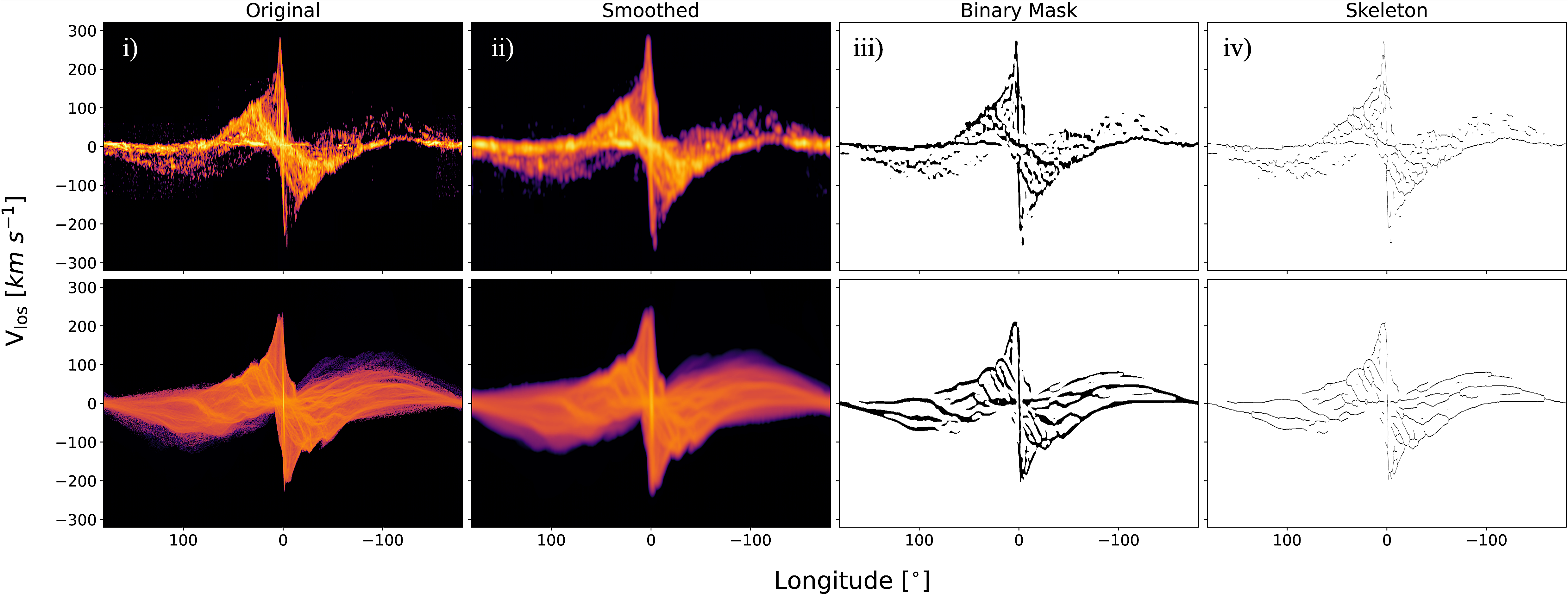}
    \caption{Different steps of the extraction of the skeletons of the observed and simulated \textit{l-v} maps. The top four panels illustrate the C0 emission map from \protect\citet{dame2001milky}, whilst the bottom panels display the example \textit{l-v} map for Model $3$ at a time $\sim 2.4$ Gyr and angle $\phi_{obs} = 32.5^{\circ}$. (i) Original image. (ii) Smoothed image with a Gaussian 2D Kernel of $\sigma_{\mathrm{l}}=1.0^{\circ}$ and $\sigma_{\mathrm{v}}=5$ km\,s$^{-1}$. (iii) Binary mask obtained after finding the peaks via a Hessian matrix and applying a binary erosion of $2$ pixels. (iv) Final skeleton image obtained by computing the medial axis of the previous image.}
    \label{fig:skeletons_obs_and_sim}
\end{figure*}

Essentially, we take the following steps to make a { skeleton} \textit{l-v} map, { also illustrated in Fig.~\ref{fig:skeletons_obs_and_sim} (for the observed CO emission on the top, and for one of our models on the bottom)}:
\begin{enumerate}
\item 
We regrid both simulations and observations to a common grid, so that the comparison can be done on a 1-to-1 basis (first column in Fig.~\ref{fig:skeletons_obs_and_sim}). We used a pixel size with $\Delta l = 0.125^{\circ}$ in the longitude axis, and $\Delta v = 1.3$ km\,s$^{-1}$ in velocity, based on the resolution of the original observations.  
\item  
We smooth these \textit{l-v} images by convolving them with a Gaussian 2D Kernel to reduce the noise, mostly to aid in highlighting the contiguous emission in the observational data along the spiral arms (rather than the ``broken'' emission from individual molecular clouds). For this convolution, we used a standard deviation of $\sigma_{l} = 1.0^{\circ}$ and $\sigma_{v} = 5$ km\,s$^{-1}$, and we do the same convolution to the modelled \textit{l-v} to keep them comparable (second column in Fig.~\ref{fig:skeletons_obs_and_sim}).  
\item 
We compute, for each pixel, the Hessian matrix and its respective eigenvalues and eigenvectors. The eigenvalues of the Hessian matrix describe the magnitude of the curvature along the direction of their respective eigenvector. Negative eigenvalues correspond to convex curvature, positive eigenvalues correspond to concave curvature, and a value of 0 is flat \citep[e.g.][]{lindeberg1996edge}. Given that the spiral arms in an \textit{l-v} map are apparent as long and thin tracks, they can be approximated to ``filamentary'' structures. We define a filament in 2D as a convex structure along the shortest axis. Therefore, for a 2D filamentary structure, at least one of the eigenvalues must be always negative, $\lambda_{1} < 0$. The second eigenvalue $\lambda_{2}$ can be either concave or convex. However, $\lambda_{1}$ should always be more strongly curved than $\lambda_{2}$. Therefore, the second requirement is that $\mid \lambda_{1} \mid > \mid \lambda_{2} \mid$. In order to avoid faint filamentary structures, we add a third requirement on the minimum amount of curvature. We set a $\lambda_{\mathrm{lim}}$ so that only regions of the map that have $\lambda_{1} < \lambda_{lim}$ are classified as filaments. This $\lambda_{lim}$ is always negative and has to be chosen manually for each different case. With the pixels that satisfy these conditions, we create a binary mask, to which we apply an extra binary erosion of $2$ pixels to thinner these wider areas (third column in Fig.~\ref{fig:skeletons_obs_and_sim}). 
\item  
We compute the medial axis of the eroded mask of the maps from (iii). This returns a structure with a width of $1$ pixel, that collects all bins containing more than one point closer to them than the boundary of the object. This is the final binary map that collects the skeleton information. A value of 1 is given to those pixels containing a skeleton, and value of 0 is given elsewhere (fourth column in Fig.~\ref{fig:skeletons_obs_and_sim}).
\end{enumerate}

We perform this process for the observations as well as for our entire set of simulations at all studied times (see Sect.~\ref{subsec: num_sampling}) and angles (see Sect.~\ref{sec:lv-plots}).

\subsection{Comparison of skeletons between simulations and observations: SMHD}
\label{Section: SMHD}

We use the Symmetrized Modified Hausdorff Distance (SMHD) statistical tool \citep[presented in][]{sormani2015recognizing} to evaluate how well our models reproduce the large scale morphology of the Milky Way. This is a symmetric version of the metric first introduced in \citet{dubuisson1994modified} and works well even with few defined parameters or contaminated data. As stated in the work by \cite{sormani2015recognizing}, this method is found to be more powerful than some of the other methods used in the literature for similar purposes, as it works well even when fewer parameters are defined or the data is contaminated.

Essentially, this technique is a comparison of pixel distances between binary images. The distance $d$ in \textit{l-v} space is defined in Eq.~\ref{eq:distance}. Given that in our case, our 2D image is not composed of two spacial axis, we need to weight the distance in each direction differently. Therefore, we set the ``uncertainty'' in the longitude axis to $\Delta l = 1^{\circ}$ (which corresponds to the angular resolution of the smoothed image we use to construct the skeletons) and in the velocity axis to $\Delta v = 10$ km\,s$^{-1}$ (which is the typical velocity width of a spiral arm). 

\begin{equation}
   \label{eq:distance}
   d(a,b) = \frac{\mid l_{a}-l_{b} \mid}{\Delta l} + \frac{\mid v_{a}-v_{b} \mid}{\Delta v}
\end{equation} 

The Modified Hausdorff Distance (MHD) is defined as the sum of the distances between each pixel $i$ containing 1 in the binary image $a$, and the closest pixel $j$ containing 1 in the binary image $b$, and the expression can be seen in Eq.~\ref{eq:MHD}. 

\begin{equation}
   \label{eq:MHD}
   \mathrm{MHD}(a,b) \equiv \sum_{i} {\rm min}_{j}(d(a_{i},b_{j}))
\end{equation}

If the set \textit{a} has \textit{N} pixels, and the set \textit{b} \textit{M} pixels, then the symmetric version is defined as seen in Eq.~\ref{eq:SMHD}. 

\begin{equation}
   \label{eq:SMHD}
   \mathrm{SMHD}(a,b) \equiv \frac{1}{2N} \mathrm{MHD}(a,b) + \frac{1}{2M} \mathrm{MHD}(b,a)
\end{equation}

Hence, lower values of this metric indicate a better fit: distances from structures in the simulations are shorter to structures in the observations (and vice versa).

Similarly to \cite{sormani2015recognizing}, we apply the SMHD metric on the skeletonised \textit{l-v}-maps, to compare the observed and model skeletons. However, note that although our procedure to extract skeletons is similar to that used by \cite{sormani2015recognizing}, it is not identical. Our method differs to that of \cite{sormani2015recognizing} in that we do not do an envelope enhancement, because we want this method to be mostly sensitive to the internal features of the galaxy (such as spiral arm tracks), rather than including the terminal velocity into the same metric. However, we then analyse the terminal velocity separately with another metric (see Sect.\,\ref{sec: terminal velocity}). We also do not do a non-maximum suppression and hystheresis thresholding, but do a medial-axis calculation instead. This, in addition to the fact that the pixel sizes and areas covered by our calculations are not the same as those in \cite{sormani2015recognizing} (therefore changing the number of “ridge” pixels in which to estimate the metric), means the values of the metric itself, as applied to our skeletonised dataset, can only be used to inter-compare our models, but cannot be directly compared to the metric values from \cite{sormani2015recognizing}.

After some experimentation, we found that this metric was prone to favouring simulations that were more crowded (i.e. had more structures/skeletons), even if such structures did not exist in the observed dataset. To quantify this effect, we include an extra metric based on the fraction of skeleton pixels that are mismatched between the two datasets. For this, we store the information of which skeleton pixel of image $a$ is matched with in image $b$ and vice versa, and then find the pairs that are uniquely matched to each other (i.e. a perfect match). All remaining pixels are considered mismatches - i.e. they represent structures that are not present on both comparison images. 

We then look at each of the individual sets: simulations (sim) and observations (obs). For each of them, we define the fraction of mismatches, $f_{\mathrm{sim}}$ or $f_{\mathrm{obs}}$, as the number of mismatched skeleton pixels ($S_{\mathrm{sim}}$ or $S_{\mathrm{obs}}$) over the total number of pixels in that skeleton ($A_{\mathrm{sim}}$ or $A_{\mathrm{obs}}$), as
\begin{equation}
    \label{eq:mismatch}
    f_{\mathrm{sim,obs}} = \frac{S_{\mathrm{sim,obs}}}{A_{\mathrm{sim,obs}}}
\end{equation}
We then add the two fractions to create a global value for the fraction of mismatches of each model, $f_{\mathrm{mis}}$, as 
\begin{equation}
    \label{eq:global}
    f_{\mathrm{mis}} = f_{\mathrm{sim}} + f_{\mathrm{obs}}
\end{equation}
\noindent that can vary between $0$ if it is a perfect match, and $2$ if there are no uniquely paired pixels. This will only act as an extra layer to compare models between each other and it is not accurate enough to rely on to predict the best model on its own. A further discussion can be seen in Appendix \ref{appendixA}.

\subsection{Terminal velocity}
\label{sec: terminal velocity}

One other way of gauging whether the models are able to reproduce the observations, is to compare the area occupied by the gas in \textit{l-v} space. To do this we need to define the outer edge of this area, i.e. the terminal velocities along a given line-of-sight. 
In order to do so for the observational data, since the edge is not nicely defined as a continuous line due to noise, we extract the polygon that roughly outlines the edges of the images by hand using the Cube Analysis and Rendering Tool for Astronomy (CARTA) \footnote{https://cartavis.org/}. 
We create the polygon mask image for the observations in  $^{12}$CO from \cite{dame2001milky} and the \ion{H}{I} from \cite{bekhti2016hi4pi} separately, we regrid the \ion{H}{I} one to match the resolution of the $^{12}$CO map, and then we add them both to create a single mask from observations. As the \textit{l-v}-maps of our simulations are noise-free and their extent is well-defined, the terminal velocity edge is defined from a simple contour of the data in \textit{l-v} space. These masks contain 1 for all pixels inside the outer edge, and 0 elsewhere. We proceed to subtract the mask of the simulations from the mask of the observations, so as to obtain an \textit{l-v} map where: matching pixels receive a value of 0, pixels only covered in observations have a value of 1, and pixels solely in the models have a value of -1. We can use this information to further infer how good a model is at reproducing the observed terminal velocities by quantifying this difference. Thus, we estimate the fraction of non-zero pixels, $m$, and average by the longitude range of the region used for the statistics, $\Delta l$, such that we obtain the average fraction of non-matched pixels per unit angle as $\langle m \rangle = m / \Delta l$. 

\section{Finding the best fit for the Milky Way structure}
\label{Section: Best Fit}

In order to compare our suite of numerical models to the observations of the Milky Way, we apply the techniques described in Sect.\,\ref{sec: galactic features} to to our simulations at a number of different timestamps. In this section we describe how we select the simulation snapshots to analyse, and the results of that analysis in terms of the best snapshot per model, as well as the overall best model at reproducing the observed galactic structure.

\subsection{Numerical sampling}
\label{subsec: num_sampling}

We analyse the galactic structure of the fifteen different models with varying stellar distributions that compose our numerical sample. We only investigate the models from a time of $\sim 2$ Gyr onwards, so that we avoid any potential structural artefacts arising from the initial settling of the disc. We study a total of $20$ snapshots per model, in intervals of $\sim 50$Myr up until the simulation reaches a total time of $\sim 3$ Gyr. For each of these snapshots, we select a sample of  6 different angles $\phi_{obs}$ around the galactic centre for the Sun position and produce the \textit{l-v} maps as explained in Sect.~\ref{sec:lv-plots}. In total, we have a sample of 1800 different simulated \textit{l-v} maps to compare to the observed $^{12}$CO emission. 

\subsection{Results from SMHD analysis}
\label{sec:SMHD_analysis}

As detailed in Section \ref{Section: SMHD}, we use the SMHD metric plus the global fraction of mismatches, $f_{\mathrm{mis}}$, for a first comparison of the galactic structures formed in the models to the CO observations from \cite{dame2001milky}. In this section, we will find the overall lowest value of SMHD in time as well as angle for each case, also dividing the galaxy in an inner region of longitude $\mid l \mid < 20^{\circ}$, and an outer region with $\mid l \mid \geq 20^{\circ}$. In doing so, we explore the effect that the spiral structures as well as the bar (if any) have on our metric and on the decision of which model provides the best fit of the Milky Way. We then proceed to perform a direct comparison of the results of the metric between the models in our numerical sample.

\subsubsection{Optimal time and angle for each model}
\label{Section: optimal time and angle}

We start by investigating how the average value of SMHD at each snapshot (averaged over all viewing angles $\phi_{obs}$) varied with time. We do not find any specific trends, which suggests that,  within the times studied, all models reached a relatively stable state in terms of the type of galactic structures formed. Therefore, we select our optimal snapshot for each model, as the time at which that model has the lowest averaged SMHD, and that is the only snapshot used for the remainder of the analysis. In order to ensure that the overall best fit was not dominated by either just the bar or the spiral arms, we further analyse this behaviour for the inner and outer Galaxy. Neither of these cases show a clear structure to the time evolution of the metric, and hence do not contribute any clear results.

 To determine the most suitable viewing angle $\phi_{obs}$ for each model at its optimal time, we look for the angle where the SMHD value is the lowest. This choice is made after ensuring that the bar's inclination relative to the observer aligns with known observations (as detailed in Sect.~\ref{sec: galactic features}). In order to quantify the possible variations of the SMHD metric value due to small changes in $\phi_{obs}$, we also produce the \textit{l-v}-plots and estimate the respective SMHD value, for two additional angles, $5^{\circ}$ apart from the original optimal $\phi_{obs}$. Our analysis confirmed that the initially selected angle was indeed the most suitable within this expanded range for all models, showing that our results are robust and not just a consequence of the limited number of angles initially considered.

Table ~\ref{Table:Results} presents the range of SMHD values for these three angles. The lowest SMHD value corresponds to the optimal $\phi_{obs}$. Additionally, Table ~\ref{Table:Results} also includes the SMHD values at the optimal time and viewing angle, as measured for the inner ($\mid l \mid < 20^{\circ}$) and outer ($\mid l \mid > 20^{\circ}$) regions of the galaxy separately.

In summary, we select the optimal time for each model based on the minimum of the angle-averaged SMHD metric, and the optimal viewing angle where the bar inclination is within observed ranges and where the SMHD metric was lowest.

\subsubsection{Optimal model}
\label{Section: Optimal model}

We now proceed to carry out a direct comparison of our models between each other, in terms of their SMHD metric. Table.~\ref{Table:Results} is a compilation of all the metrics we use to inter-compare models.  We include the range of values for the global SMHD metric, fraction of mismatches and terminal velocity metric at the optimal time for the optimal viewing angle $\phi_{obs} \pm 5^{\circ}$. For all cases, the lowest value correspond to the final chosen viewing angle. We also include the inner and outer values of this metrics, as well as the bar parameters (half-length $L/2$, inclination $\alpha$, and pattern speed $\Omega$ , as determined in Section \ref{sec:BAR}).

\begin{figure*}
	\includegraphics[width=0.98\columnwidth]{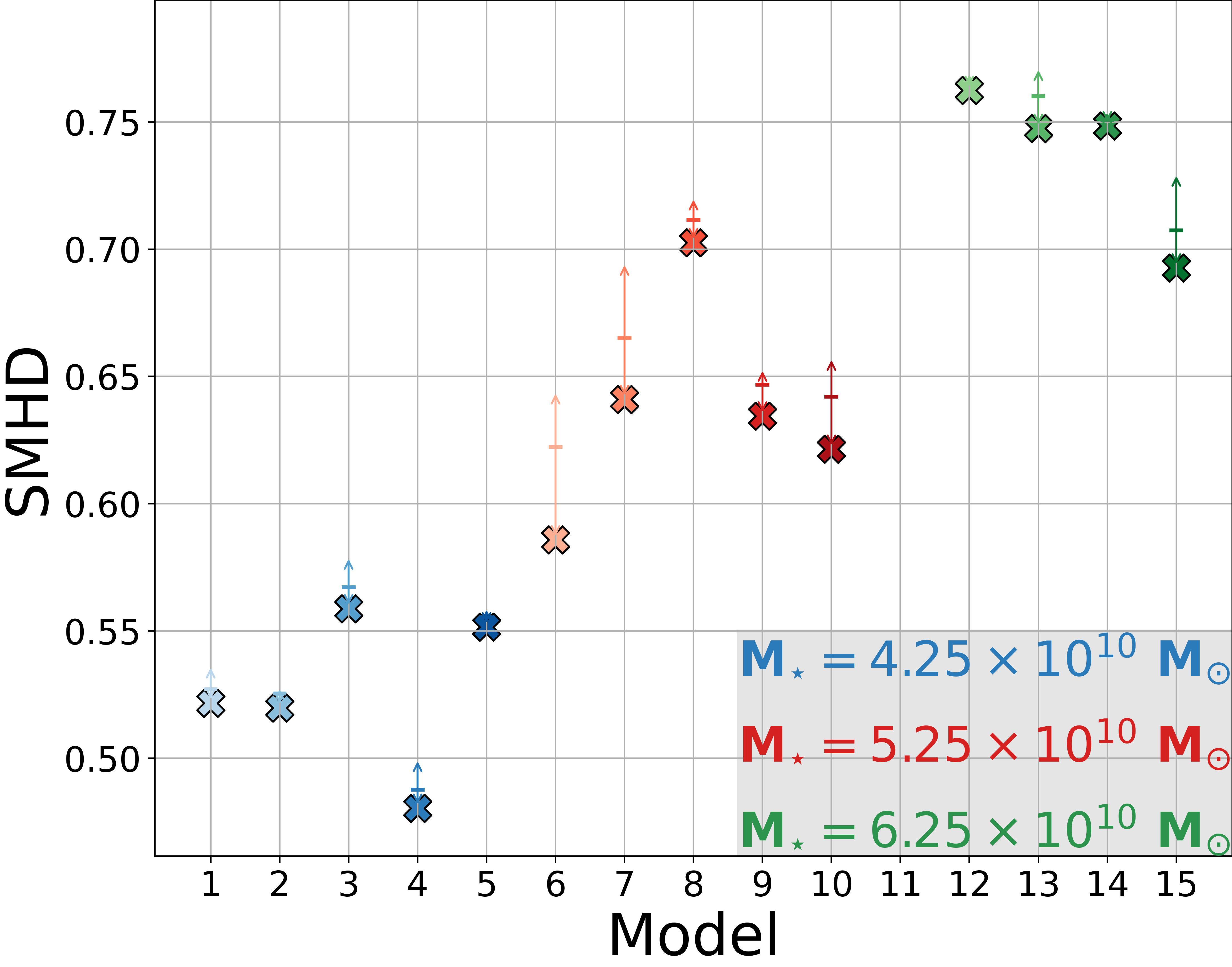}
    \hfill
 	\includegraphics[width=\columnwidth]{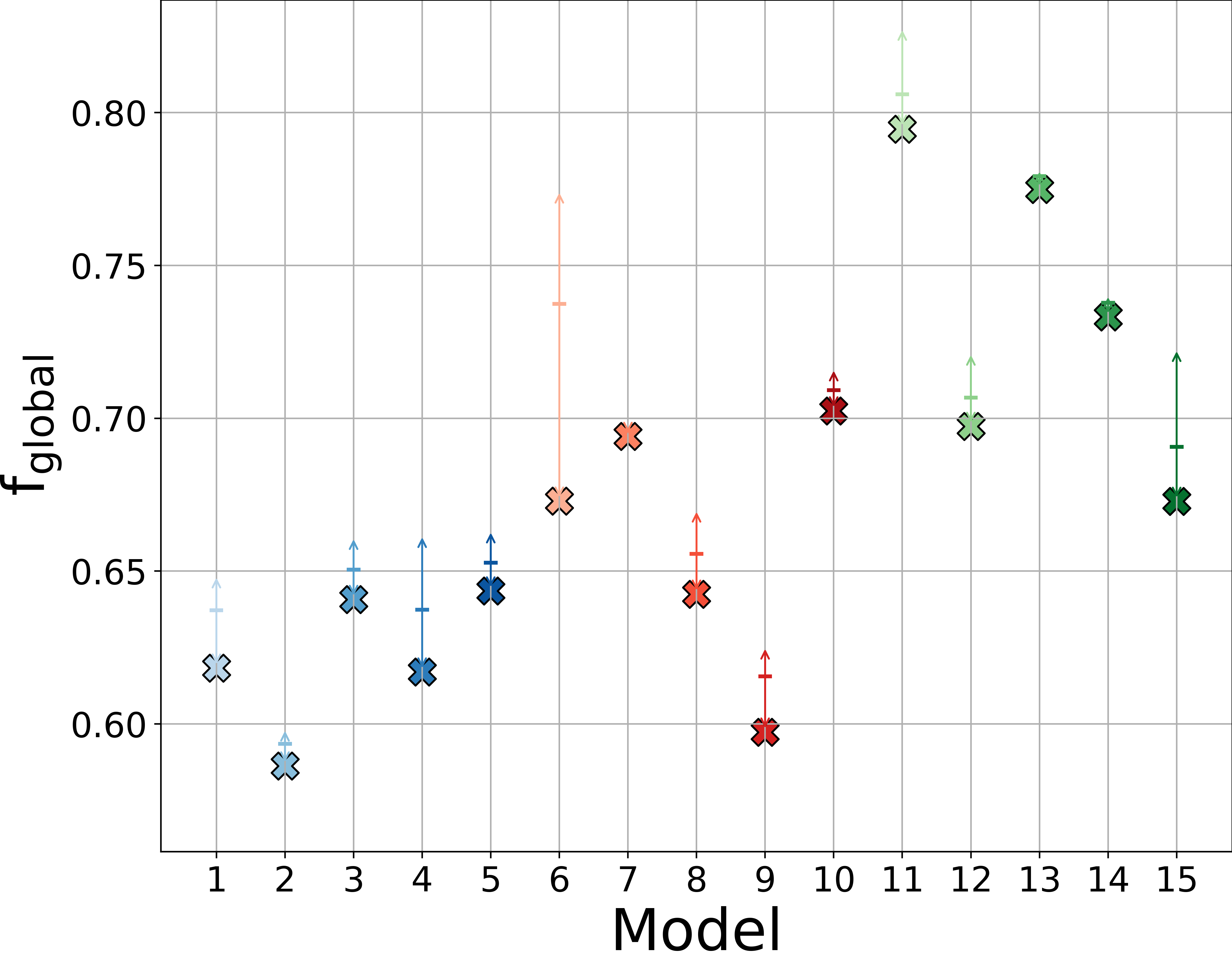}
    \caption{Variation of the final SMHD value (left) and global fraction of mismatches (right) for the 15 models explored, at their selected time and viewing angle (from Section.~\ref{Section: optimal time and angle}). For both panels, the data points have been colour-coded so that for a fixed initial stellar mass, lighter to darker colours indicate a decrease in the initial bulge fraction: $25 \%$, $20 \%$, $15 \%$, $10 \%$ and $5 \%$. Blue data points indicate models with an initial stellar mass of M$_{\star} = 4.25 \times 10^{10}$ M$_{\odot}$, whilst red indicates M$_{\star} = 5.25 \times 10^{10}$ M$_{\odot}$ and green M$_{\star} = 6.25 \times 10^{10}$ M$_{\odot}$. The arrows indicate the variation of SMHD within each model for the optimal viewing angle $\phi_{obs} \pm 5^{\circ}$, with the median marked as a horizontal line. }
    \label{fig:SMHD_model}
\end{figure*}

In order to see how the models compare to each other, we represent the variation of the final SMHD against the models in Fig.~\ref{fig:SMHD_model}. The data points have been colour-coded so that for a fixed initial stellar mass, lighter to darker colours indicate a decrease in the initial bulge fraction: $25 \%$, $20 \%$, $15 \%$, $10 \%$ and $5 \%$. Blue data points indicate models with an initial stellar mass of M$_{\star} = 4.25 \times 10^{10}$ M$_{\odot}$, whereas red indicates M$_{\star} = 5.25 \times 10^{10}$ M$_{\odot}$ and green M$_{\star} = 6.25 \times 10^{10}$ M$_{\odot}$.  We have added an arrow to indicate the variation of the SMHD with the range of $\pm 5^{\circ}$, as well as a horizontal line at the median of the distribution, for easier comparison of the variability with viewing angle between models. It is clearly visible that the lower values for the metric are achieved with the lower initial stellar mass. Therefore, we exclude all models above $5$ as the SMHD metric keeps increasing. Model $1$ can also be dismissed, as it does not generate a galactic bar.

\begin{center}

\begin{table*}
    \centering
    \begin{tabular}{ *{14}{c|} }
        \hline
         Model & Time & SMHD & SMHD$_{\mathrm{inn}}$ & SMHD$_{\mathrm{out}}$ & $f_{\mathrm{global}}$ & $f_{\mathrm{inn}}$ & $f_{\mathrm{out}}$ & $\langle m_{\mathrm{global}} \rangle$ & $\langle m_{\mathrm{inn}} \rangle$ & $\langle m_{\mathrm{out}} \rangle$ & $L/2$ & $\alpha $ & $\Omega $ \\
        \hline
        1 & 2400 & [0.52, 0.54] & 0.79 & 0.44 & [0.62, 0.65] & 0.99 & 0.54 & [292, 299] & 680 & 249 & nan & 32   & 21.0 $\pm$ 2.0 \\
        2 & 2700 & [0.52, 0.53] & 0.71 & 0.46 & [0.59, 0.60] & 0.97 & 0.48 & [250, 252] & 518 & 217 & 3.8 $\pm$ 0.8 & 45   & 32.0 $\pm$ 2.0 \\
        3 & 2600 & [0.56, 0.58] & 0.83 & 0.48 & [0.64, 0.66] & 1.01 & 0.54 & [261, 265] & 503 & 235 & 3.1 $\pm$ 0.8 & 20   & 38.2 $\pm$ 0.4 \\
        4 & 2650 & [0.48, 0.50] & 0.61 & 0.44 & [0.62, 0.66] & 0.88 & 0.53 & [223, 255] & 457 & 219 & 3.2 $\pm$ 0.8 & 45   & 30.0 $\pm$ 0.2 \\
        5 & 2050 & [0.55, 0.56] & 0.77 & 0.49 & [0.64, 0.66] & 0.99 & 0.53 & [228, 230] & 548 & 189 & 3.4 $\pm$ 0.8 & 32   & 31.2 $\pm$ 0.7 \\
        6 & 2100 & [0.59, 0.64] & 0.9 & 0.49 & [0.67, 0.77] & 1.11 & 0.54 & [274, 321] & 481 & 248 & 4.0 $\pm$ 0.8 & 20   & 29.2 $\pm$ 0.1 \\
        7 & 2450 & [0.64, 0.70] & 1.04 & 0.54 & [0.69, 0.70] & 1.13 & 0.57 & [216, 217] & 543 & 175 & 2.8 $\pm$ 0.8 & 20   & 27.3 $\pm$ 0.3 \\
        8 & 2150 & [0.70, 0.72] & 1.17 & 0.56 & [0.64, 0.67] & 1.08 & 0.53 & [246, 254] & 401 & 227 & 4.9 $\pm$ 0.8 & 45   & 26.8 $\pm$ 0.2 \\
        9 & 2100 & [0.63, 0.65] & 0.96 & 0.54 & [0.60, 0.63] & 0.98 & 0.51 & [221, 223] & 272 & 215 & 3.1 $\pm$ 0.8 & 20   & 26.1 $\pm$ 0.2 \\
        10 & 2200 & [0.62, 0.66] & 1.07 & 0.48 & [0.70, 0.72] & 1.09 & 0.58 & [216, 248] & 445 & 199 & 3.7 $\pm$ 0.8 & 45   & 26.7 $\pm$ 0.2 \\
        11 & 2950 & [0.83, 0.84] & 1.4 & 0.66 & [0.79, 0.83] & 1.23 & 0.67 & [273, 282] & 822 & 206 & 4.4 $\pm$ 0.8 & 45   & 29.0 $\pm$ 0.2 \\
        12 & 2050 & [0.76, 0.77] & 1.44 & 0.55 & [0.70, 0.72] & 1.11 & 0.58 & [307, 319] & 728 & 260 & 3.6 $\pm$ 0.8 & 32   & 29.0 $\pm$ 0.2 \\
        13 & 2100 & [0.75, 0.77] & 1.16 & 0.62 & [0.77, 0.78] & 1.12 & 0.67 & [264, 288] & 663 & 234 & 2.8 $\pm$ 0.8 & 45   & 26.5 $\pm$ 0.4 \\
        14 & 2050 & [0.75, 0.76] & 1.22 & 0.61 & [0.73, 0.74] & 1.17 & 0.6 & [271, 284] & 448 & 258 & 4.0 $\pm$ 0.8 & 20   & 20.8 $\pm$ 0.8 \\
        15 & 2350 & [0.69, 0.73] & 1.22 & 0.54 & [0.67, 0.72] & 1.09 & 0.55 & [272, 855] & 684 & 221 & 4.1 $\pm$ 0.8 & 20   & 26.8 $\pm$ 0.2 \\

        \hline
    \end{tabular}
    \caption{Table containing results for the best time and viewing angle for all different measurements and metrics performed on the fifteen different models that compose our numerical sample. From left to right: { Model number; optimal time; SMHD metric for global, inner and outer galaxy (SMHD, SMHD$_{\mathrm{inn}}$ and SMHD$_{\mathrm{out}}$); global, inner and outer fraction of mismatches ($f_{\mathrm{global}}$, $f_{\mathrm{inn}}$, and $f_{\mathrm{out}}$); global, inner and outer angle averaged terminal velocity metric ($\langle m_{\mathrm{global}} \rangle$, $\langle m_{\mathrm{inn}} \rangle$ and $\langle m_{\mathrm{out}} \rangle$); bar half-length ($L/2$); bar inclination ($\alpha $); and bar pattern speed $\Omega $. 
     }}
    \label{Table:Results}
\end{table*}

\end{center}

In order to assess how good a match any given model is at its optimal time and viewing angle, we not only check that the SMHD metric is low, but also that the number of mismatches is low. Hence we then investigate the fraction of mismatches and check whether they concur with the SMHD metric. For all models $f_{\mathrm{inn}}$ is always higher than $f_{\mathrm{out}}$ (as seen in Table.~\ref{Table:Results}), suggesting that there are more unmatched structures towards the inner galaxy. The global fraction of mismatches for each model can be seen in the right panel of Fig.~\ref{fig:SMHD_model}. Similarly to the SMHD, the value of $f_{\mathrm{global}}$ generally increases with increasing initial stellar masses. There is one clear exception to this trend, for model $9$. This model has values of $f_{\mathrm{global}}$, $f_{\mathrm{inn}}$ and $f_{\mathrm{out}}$ similar to those of models 2-5. However, given that its global SMHD is significantly worse than models 2-5, we do not consider this model further.

In our comparison of models with a bar and the lowest global SMHD values (i.e. models $2$, $3$, $4$ and $5$), we find that the fraction of mismatches is relatively similar across these models, both globally and within the inner/outer regions of the Galaxy. Notably, models 2 and 4 exhibit the lowest mismatch fractions. Among them, model 2 has a marginally lower overall fraction of mismatches. However, model 4 demonstrates a reduced fraction of mismatches in the inner Galaxy ($f_{\mathrm{inn}}$), suggesting it more accurately represents the inner galactic structures. Furthermore, the global SMHD metric for model 4 is distinctly lower than that of model 2, including both the inner and outer values. These findings indicate that model 4 is a better overall match of the structures in $l-v$ space.

\begin{figure*}
	\includegraphics[width=\textwidth]{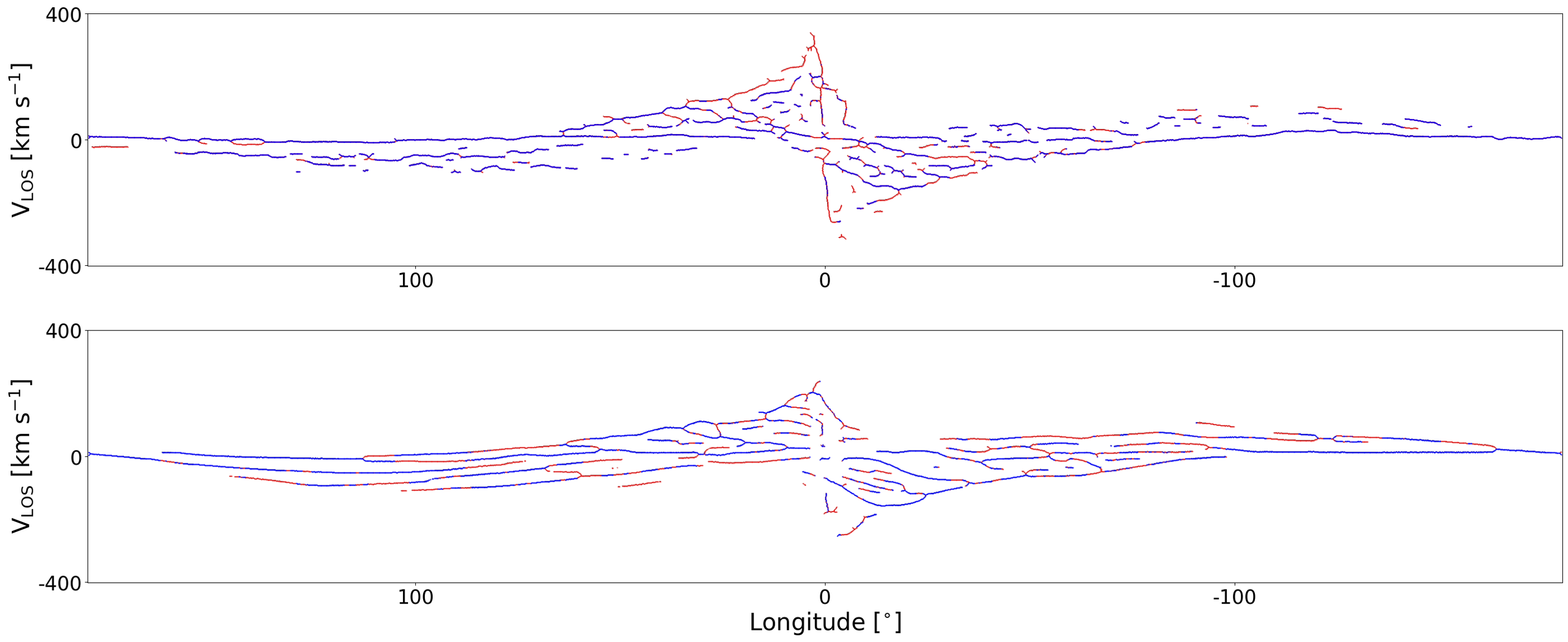}
    \caption{ Skeletons extracted from the \textit{l-v} maps of the $^{12}$CO observations by \citet{dame1986largest} (top) and our Model 4 at a time $\sim 2.6$ Gyr and angle $\phi_{obs} = 45^{\circ}$ (bottom). Blue pixels are uniquely associated with each other, whilst red-coloured ones display those defined as mismatches.}
    \label{fig:frac_mism}
\end{figure*}

Fig.~\ref{fig:frac_mism} illustrates this point further. It displays the extracted skeletons from the l-v maps of the $^{12}$CO observations by \citet{dame1986largest} at the top, and our Model 4 at approximately $\sim 2.6$ Gyr with an observer angle $\phi_{obs} = 45^{\circ}$ at the bottom. The pixels are color-coded, with blue representing uniquely associated pixels and red indicating those defined as mismatches. This figure reveals that, while most structures in the simulation’s inner and outer regions correspond to those found in the observations, Model 4 struggles to replicate the highest velocities observed in the inner Galaxy.

Although no model is a perfect match to the observations, the SMHD metric and the fraction of mismatches collectively suggest Model 4 as the most promising candidate, with Model 2 closely following. The variability of SMHD values with small changes in viewing angle is minimal, as indicated by the arrows in the left plot of Fig.~\ref{fig:SMHD_model}.  We further investigate if the selection of a different bar inclination would affect the final best model chosen by the SMHD metric, and we find that Model 4 still has an overall lower value when exclusively looking at inclinations of $20^{\circ}$ and $32.5^{\circ}$. Similarly, the variation in the fraction of mismatches   with viewing angle is not significantly large, as observed in the right plot of the same figure. Nevertheless, to further refine our assessment of which model best represents the structure of the Milky Way, we will also incorporate the terminal velocity as an additional evaluative measure.

\subsection{Comparison of the terminal velocity}
\label{results:term-vel}

\begin{figure*}
	\includegraphics[width=\textwidth]{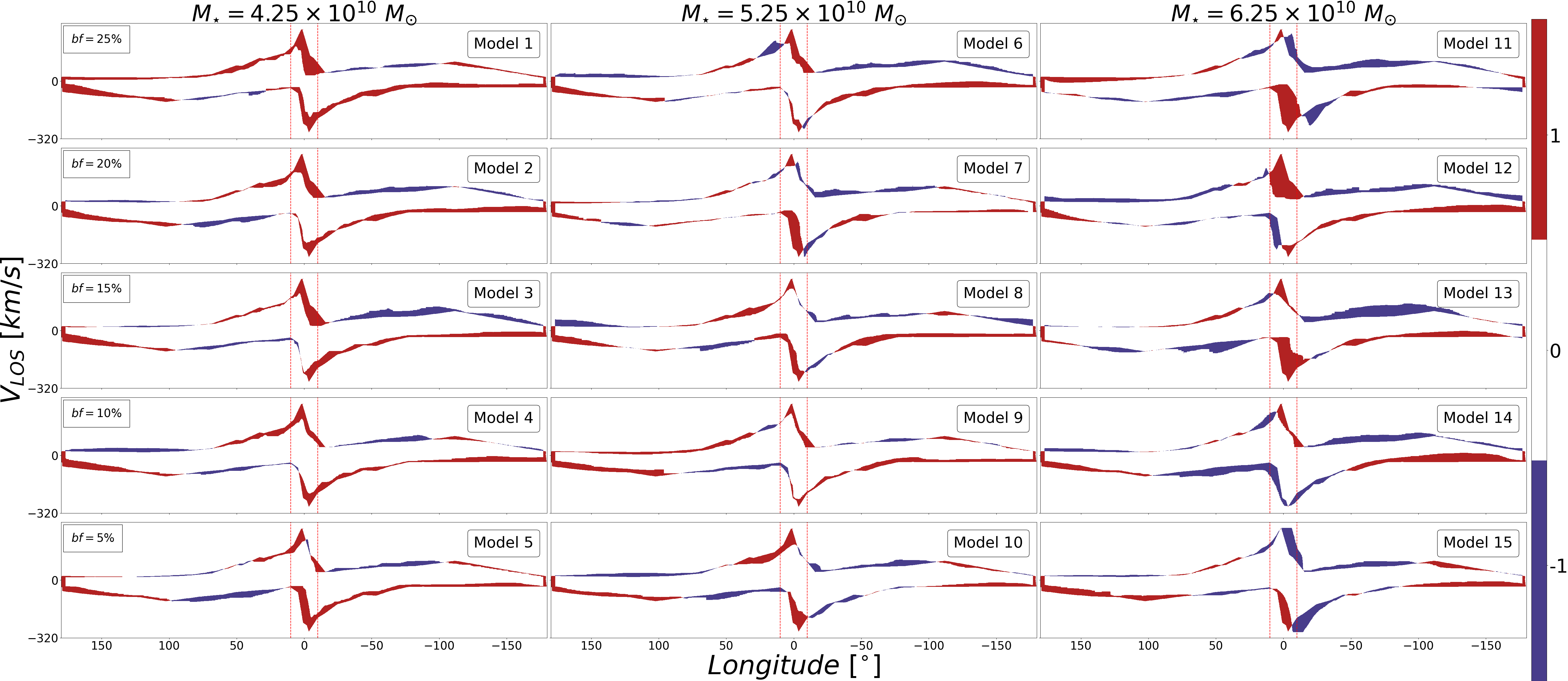}
    \caption{Longitude-velocity diagrams showing the comparison of the terminal velocity of the \citet{dame2001milky} $^{12}$CO observations combined with \ion{H}{I} data from \citet{bekhti2016hi4pi} with our fifteen different models. Columns increase in initial stellar mass from left to right, and rows decrease in bulge fraction from top to bottom. Red indicates pixels present in the observations but not in the simulations, where blue refers to those present in the simulations, but not on the observations. The vertical dashed red lines divide the plots in the inner ($\mid l \mid < 20^{\circ}$) and outer ($\mid l \mid < 20^{\circ}$) regions.}
    \label{fig:outer_edge}
\end{figure*}

Besides the imprints of the spiral/bar pattern in the \textit{l-v} maps, one other test to deduce how well our models are able to reproduce the observed velocity pattern of the Galaxy, is to compare the terminal velocities that our models reach, versus the observed ones. We do this test solely based on the selected best overall time and viewing angle for each of our models as per the results from the SMHD metric (from Sect.\,\ref{Section: optimal time and angle}).
As explained in Sect.\,\ref{sec: terminal velocity}, we obtain the outer edge of the area occupied by the gas in the \textit{l-v} maps, and calculate the difference between the area covered by the models, and that of the observations. Once more, we divide this analysis in overall, inner and outer galaxy to gauge the weight of the different areas on the overall result. 

The results from this analysis are shown in Fig.~\ref{fig:outer_edge}. Here, red pixels indicate those regions from the observed $^{12}$CO and \ion{H}{I} emission in the \textit{l-v} map that are not present in the \textit{l-v} space of the simulations. On the other side, blue pixels specify the regions in \textit{l-v} of our models that do not appear on the observations. White pixels inside the contours represent the velocity-longitude pairs that appear in both observations and simulations. The division between inner and outer galaxy is shown with the two vertical red dashed lines. A visual comparison of the results can be achieved by looking at those non-white pixels in Fig.~\ref{fig:outer_edge}. The bigger the red/blue areas in the individual images, the more discrepancies that model exhibits with the observations. It is possible to see that models with larger stellar masses tend to have a larger number of non-zero pixels, making these candidates less suitable to reproduce the observations. This result is in accordance with the outcomes from the SMHD metric. A further discussion can be seen in Appendix~\ref{appendixC}.

As explained in Sect.\,\ref{sec: terminal velocity}, to quantify this difference, we estimate the angle-averaged number of non-zero pixels, $\langle m \rangle = m / \Delta l$, where $\Delta l$ is the longitude range of the region used for the statistics. The angle-averaged number of non-zero pixels for the overall, inner and outer galaxy,$\langle m_{\mathrm{global}} \rangle$, $\langle m_{\mathrm{inn}} \rangle$ and $\langle m_{\mathrm{out}} \rangle$, are collated in Table \ref{Table:Results}.  Similarly to the results from the SMHD metric, the models with the lowest $\langle m_{\mathrm{global}} \rangle$ are models 2, 3, 4 and 5. Focusing on the preferred models from the SMHD analysis in Section.~\ref{Section: Optimal model} (models $2$ and $4$),  Model $4$ presents the lowest values in the overall $\langle m \rangle$ metric, as well as in the inner and outer regions of the galaxy, meaning that it is a better reproduction of the Galaxy velocity profile. Therefore, we conclude that out of our suite of 15 models,  Model $4$ is consistently the best at reproducing the studied observations of the Milky Way, showing the lowest values of the terminal velocity and SMHD metric. We adopt this as our best model and proceed to further analyse how some specific galactic features compare to those of the Milky Way.

\section{Characterisation of the best model }
\label{Section: Properties}
\subsection{The Galactic Bar}
\label{sec:BAR}

As mentioned in Section~\ref{sec:intro}, there is some debate on the different observational  traits of the galactic bar. In this section we investigate the bar length measured for the models in our sample, with particular focus on the values obtained for our best model. Note, however, that we do this in a simplistic way, as our intent is purely to have an estimate of the overall length, orientation, and pattern speed of the bars that are formed in the models as seen in the gas. Fitting the stellar bar more accurately is beyond the scope of this paper, but on Appendix.~\ref{appendixD}, we show the 3D distribution of the stellar bar, which suggests that it has (at least) two components - a longer thinner bar, and a boxy/peanut bulge-like structure at the centre. In Appendix~\ref{appendixD} we also show the radial and circular components of the velocity of the stars, showing a very similar pattern to that seen in the gas (see Sect.\ref{sec:gal_dynamics}).   For consistency, from here onwards we will use the half-length of the bar (3-5 kpc) as a reference for our analyses (see Section~\ref{sec:intro} for further information).

For the purpose of this paper, we will simply use the gas top-down surface density distribution (where the transitions are sharper, see figures from Appendix~\ref{appendixA}), and we use the CARTA software to manually define the bar ends  (i.e. the position where the linear bar meets the beginning of the spiral arms) from those top-down gas surface density maps.  
Table~\ref{Table:Results} summarises the results, where    errors on the length were obtained via a Monte Carlo error propagation, assuming an initial uncertainty of $500$\,pc in the definition of the bar ends. The bar half-length ($L/2$) is defined as half the distance between the extremities, and ranges between 2.8 and 4.9\,kpc in our   barred models. Model 4, our best model, generates a bar with a half-length of $3.1 \pm 0.8$\,kpc, meaning that the bar extends to $\sim 3$\,kpc from the galactic centre. This result is in agreement with recent studies that use data from \textit{Gaia} EDR3, such as \cite{queiroz2021milky} or \cite{lucey2023dynamically}, which find, using $6$D phase-space information, that the Galactic bar extends to about  $\sim 4$\,kpc from the Galactic Centre.

To determine the pattern speed of the bars in our simulations, we analyze the change in angles derived from the WPCA analysis (see Section.~\ref{sec:lv-plots}) at three distinct times: the optimal time identified via the SMHD metric, and two additional times, at $50$\,Myr before and 50\,Myr after the optimal time. This approach provides us with three measurements for the pattern speed. 
The final pattern speed is calculated as the mean of these values. The associated error is estimated based on the standard deviation of these measurements. The results for the pattern speed ($\Omega$) can be seen in Table~\ref{Table:Results}, where these values range between $20.8 \pm 0.8$\,$\mathrm{km}\ \mathrm{s}^{-1}\ \mathrm{kpc}^{-1}$ to $38.2 \pm 0.4$\,$\mathrm{km}\ \mathrm{s}^{-1}$, with a value of $30.0 \pm 0.2$ $\mathrm{km}\ \mathrm{s}^{-1}\ \mathrm{kpc}^{-1}$ for our best model (Model 4). Values deduced from observations are in between $30 - 40$ $\mathrm{km}\ \mathrm{s}^{-1}\ \mathrm{kpc}^{-1}$ \citep[e.g.][]{drimmel2023,clarke2022}, thus our best model is within the observed range.

\subsection{The Spiral Pattern}
\label{sec:spiral pattern}

\begin{figure*}
	\includegraphics[width=0.9\textwidth]{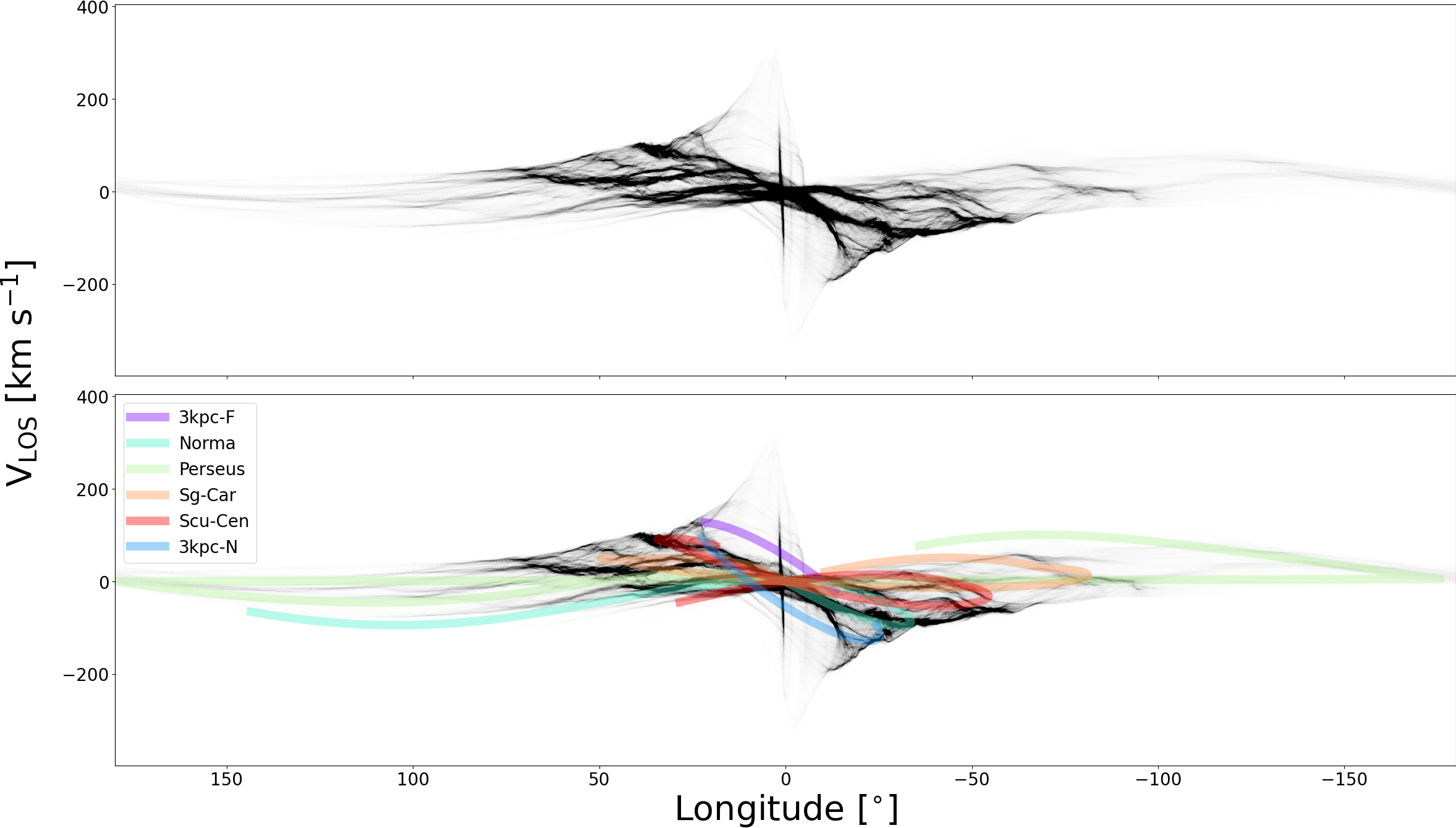}
    \caption{Top: Longitude-velocity map for our best model,  Model $4$, at a viewing angle  $\phi_{obs} = 45 ^{\circ}$. Bottom: Same image, where the spiral arms tracks from \citet{taylor1993pulsar} are superimposed. These tracks are displayed with a width of 10\,$\mathrm{km}\ \mathrm{s}^{-1}$ and the different colours represent different spiral arms: the 3\,kpc-arm is represented in blue for the Near and purple for the Far counterparts; and light blue, light green, yellow and orange represent the Norma-Outer, Perseus, Sagittarius-Carina and Scutum-Centaurus arms.  }
    \label{fig:Spiral Pattern -LV}
\end{figure*}

\begin{figure*}
        \includegraphics[width=\textwidth]{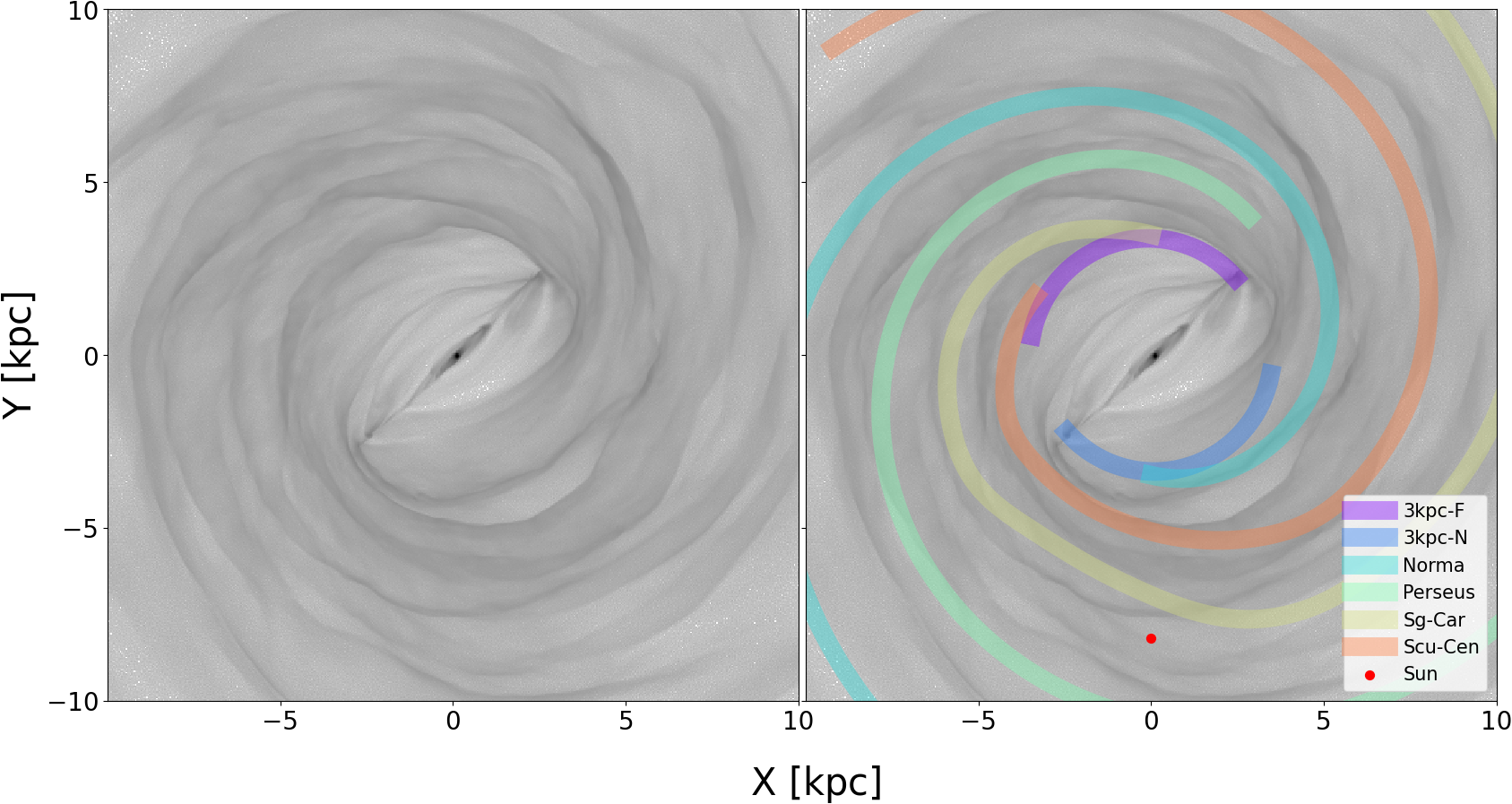}
    \caption{Same as Fig.~\ref{fig:Spiral Pattern -LV} for the $20 \times 20$\,kpc$^{2}$ face-on view of our best model,  Model $4$. Original image is shown on the left, whilst the spiral tracks from \citet{taylor1993pulsar} are superimposed on the right panel. The Sun was positioned at a distance from the Galactic Centre of 8.2\,kpc and represented with a red dot.}
    \label{fig:Spiral Pattern -XY}
\end{figure*}

As a means to visually compare the spiral pattern formed in our best model, to some of the most commonly used spiral arm tracks of the Milky Way, we overplot on our best model, the spiral track models by \cite{taylor1993pulsar} updated by \cite{cordes2004erratum}, which are used by several Milky Way molecular gas surveys \citep[e.g.][]{colombo2022sedigism}. We do so both in the \textit{l-v} diagram as well as the top-down view of the Galaxy. The \textit{l-v} map at the best viewing angle of $\phi_{obs} = 45^{\circ}$ and the $20 \times 20$ kpc$^{2}$ face-on view of Model 4 can be seen in Fig.~\ref{fig:Spiral Pattern -LV} and Fig.~\ref{fig:Spiral Pattern -XY}, both with and without the spiral tracks from \cite{taylor1993pulsar} superimposed in colour. From these plots, we can see that the spiral arm loops (in \textit{l-v} space) not always match exactly the position of the loops in the model which could be partly a consequence of us sampling the viewing angle only every 12.5 degrees. 

The spiral tracks from \cite{taylor1993pulsar} are also idealised log spirals with specific pitch angles (with the exception of the kinks closer to the Sun), which is also not always a perfect match to the pattern seen in molecular gas tracers. Indeed, perfect log-spirals are not reproduced in our live models, as our spiral pattern is more transient and dynamic in nature. Furthermore, the specific choice of spiral arms from different observational works would place the tracks in  slightly different positions \citep[e.g.][]{reid19}. 
Similarly to the \textit{l-v} space, we can see that some spiral arm tracks in the top-down view are well matched, while others have some small shifts with respect to the position of our spiral arms. Nevertheless, we note that the top-down projection of these tracks from \cite{taylor1993pulsar} are highly uncertain,  thus should only be taken as purely illustrative.

Nevertheless, even though we do not try to  match our models to these idealised tracks (as we do the comparison directly on the observational data, without assuming any specific model of the spiral pattern), we can see that in general our best model resembles the  idealised global structure of the Milky Way.

\subsection{Galactic Dynamics}
\label{sec:gal_dynamics}

\begin{figure*}
	\includegraphics[width=\textwidth]{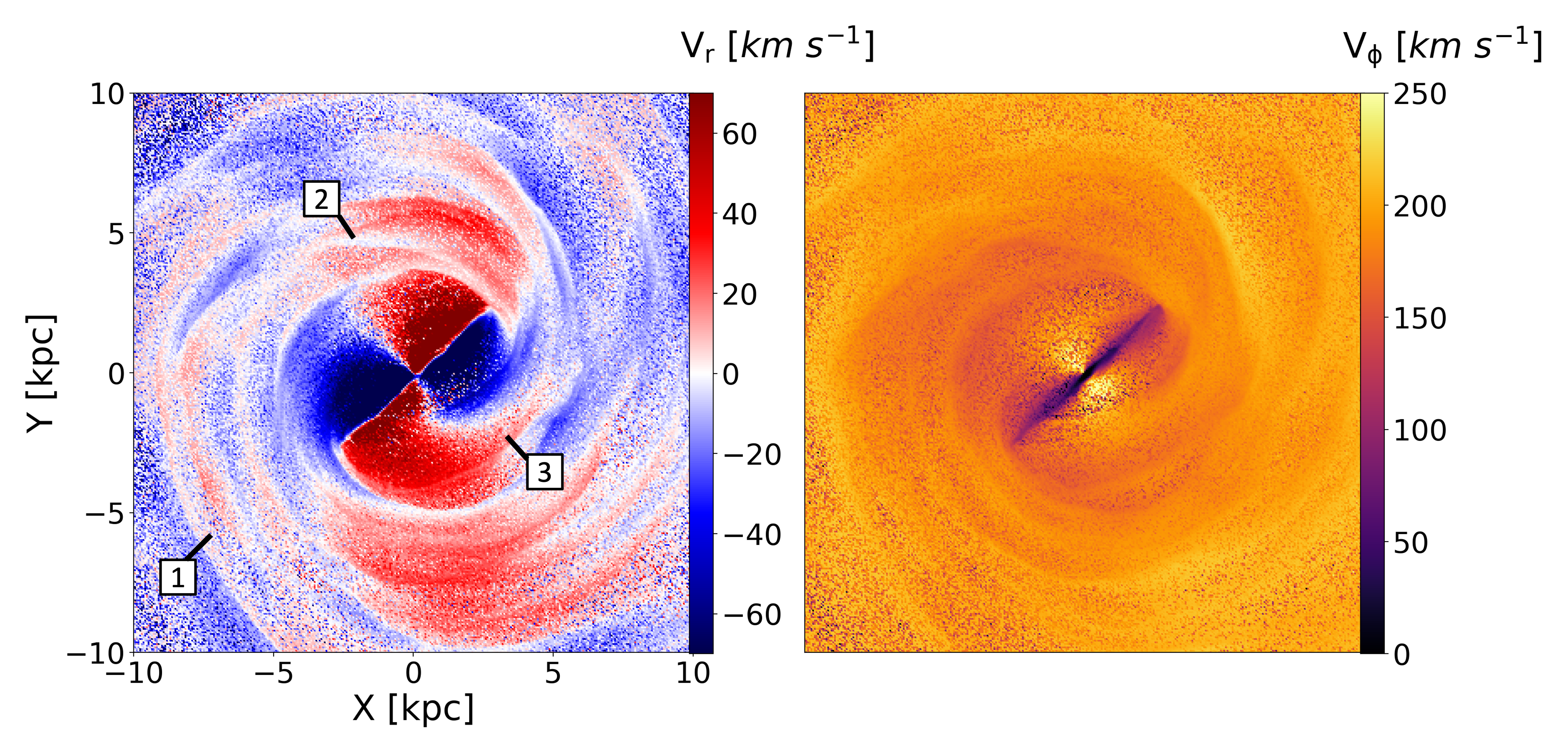}
    \caption{Top-down view of the $20 \times 20$ kpc$^{2}$ faded surface density of  Model $4$. The left panel is colour-coded with the radial velocity, V$_\mathrm{rad}$, of the gas cells, whilst the right panel shows the tangential component, V$_{\phi}$. Black solid lines in the former represent the cross-section of the spiral arms analysed in this model. }
    \label{fig:Velocitiesd-butterfly}
\end{figure*}

We investigate the dynamics generated by our best model by looking at the radial and tangential (circular) velocity components of the gas in our  Model $4$, V$_{\mathrm{rad}}$ and V$_{\mathrm{tan}}$.  Note that in Appendix~\ref{appendixD} we also show the radial and tangential velocity components for the stars, showing a very similar pattern to that seen in the gas, albeit with less sharp changes . The top-down view of the simulation with the respective values of the gas velocities can be seen in Fig.~\ref{fig:Velocitiesd-butterfly}, where the surface density of the gas has been faded in the background for a visual comparison with the positions of the spiral arms and bar. The solid black lines with numbers represent the cross-section of the spiral arms in the sample that we investigate further down.

From the left panel of Fig.~\ref{fig:Velocitiesd-butterfly}, we can see that the galactic centre presents a ``butterfly'' or quadrupole pattern in radial velocity, typical of barred-galaxies \citep[e.g.][]{buta2001dust,buta2004distribution,cuomo2021bar,querejeta2016gravitational}. There is a four-quadrant section delimited by the symmetry axes of the galactic bar, where the radial velocities of the gas cells shift from positive to negative values. This feature has directly been observed in the Milky Way using \textit{Gaia} measurements  
\citep[e.g.][]{queiroz2021milky,leung2022}. It is also possible to see that the spiral arms are mostly located in regions where there is a drastic change in V$_\mathrm{rad}$, creating shock fronts where the gas accumulates and creates the spiral arm overdensities, and the gas in the inter-arm regions is mostly radially static, with very low V$_\mathrm{rad}$.

From the right panel of Fig.~\ref{fig:Velocitiesd-butterfly}, we can see how the tangential velocity of the gas varies in our simulation. The galactic bar in this case produces very low values of V$_{\phi}$ along its major axis, as this is where the gas is mostly moving radially. The well-defined spiral arms also show lower values of V$_{\phi}$ (thus contributing to reduced shear), whilst the gas in the inter-arm regions have higher V$_{\phi}$ and thus induce larger shear motions from the differential rotation.

\begin{figure}
	\includegraphics[width=\columnwidth]{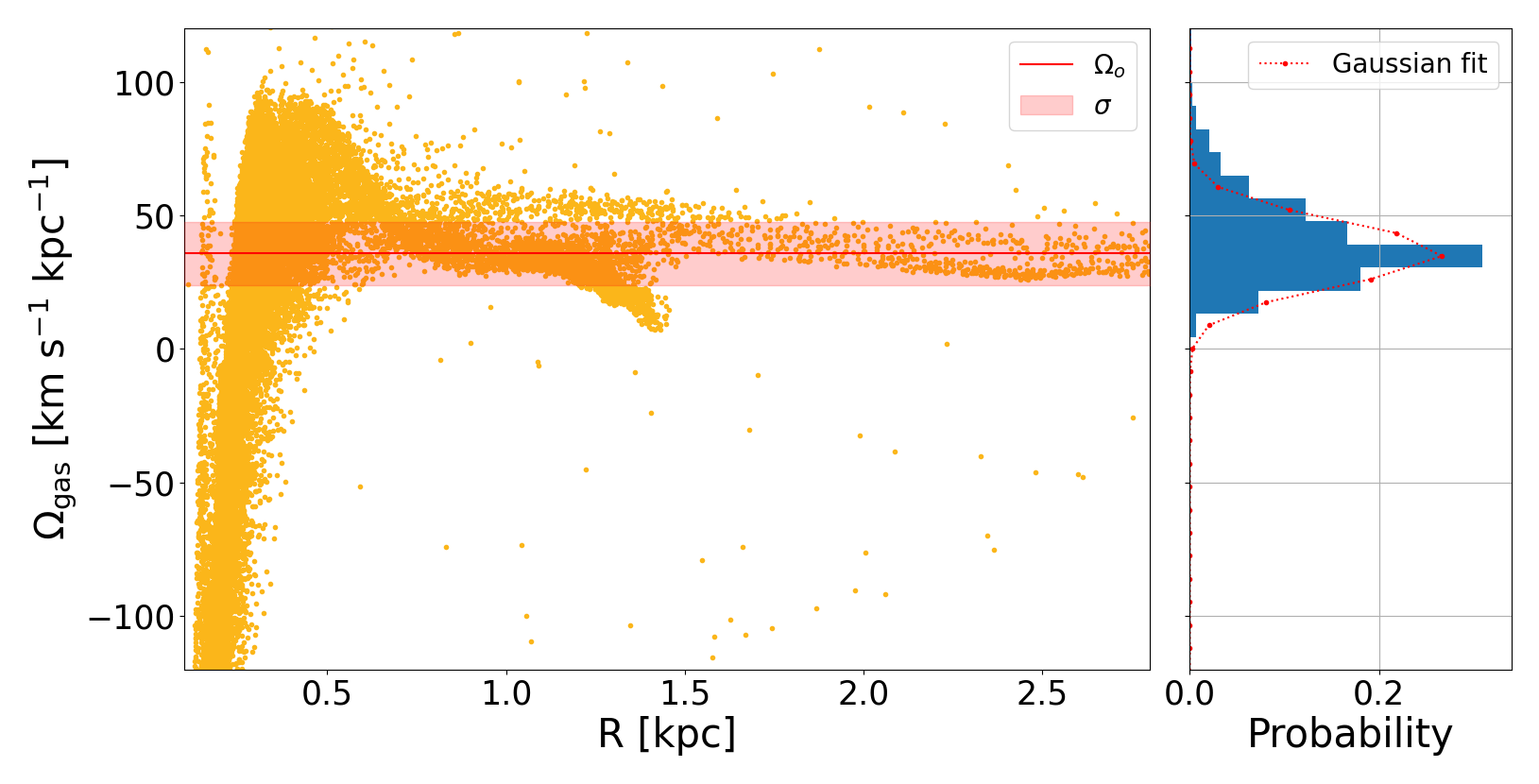}
    \caption{ Left: $\Omega_{\mathrm{gas}}$ against distance from the Galactic Center (R) for those gas cells that fall within 5 pixels (or $200$\,pc) from the bar major axis. Here, $\Omega_{\mathrm{gas}}$ is defined as $\mathrm{V}_\mathrm{tan}(\mathrm{R})/\mathrm{R}$. Solid red line indicates the fitted peak of the $\Omega_{\mathrm{gas}}$ distribution using a Gaussian least-square fitting for gas cells located at $R>0.5$\,kpc ($\Omega_{o}$). The red shaded area represents the dispersion of the data ($\sigma$). Right: Vertical histogram of the data for $\Omega_{\mathrm{gas}}$, with the respective gaussian fit overlayed as the dotted red line.} 
    \label{fig:Pattern Speed}
\end{figure}

Furthermore, we look at how the tangential  (circular) velocity of the gas  along the bar major axis vary with radial distance from the Galactic Centre ($R$),  by taking the velocities of all gas cells within $200$\,pc from the bar major axis.  Therefore, we define a new quantity $\Omega_{\mathrm{gas}}$ as $\mathrm{V}_\mathrm{tan}(R)/R$ with dimensions of pattern speed, and we represent it against $R$ in the left panel of Fig.~\ref{fig:Pattern Speed}. We can see that outside the central-most region (beyond $R>0.5$\,kpc), the circular velocity of the gas becomes constant and with extremely low dispersion, suggesting that the gas along the bar has near-constant angular velocity, thus mimicking a solid body-type of rotation. We can estimate the mean value and its dispersion ($\sigma$) by creating a histogram of the data and performing a Gaussian least-square fitting.  The resulting value for distances larger than $R>0.5$\,kpc indicate 
$\Omega_{\mathrm{gas}}$\,=\,35.8\,$\pm$\,11.8\,$\mathrm{km}\ \mathrm{s}^{-1}\ \mathrm{kpc}^{-1}$. The solid red line on the left panel of Fig.~\ref{fig:Pattern Speed} indicates the fitted peak of the $\Omega_{\mathrm{gas}}$ distribution, whilst the red shaded area represents the dispersion of the data ($\sigma$). The right panel of Fig.~\ref{fig:Pattern Speed} shows the vertical histogram of the data for $\Omega_{\mathrm{gas}}$, with the respective gaussian fit overlayed as the dotted red line. The resulted value for $\Omega_{o}$ is comparable with the values found for the bar pattern speed (see Table.~\ref{Table:Results}), suggesting that as the gas travels across the bar major axis,  its trajectory changes, affected by the potential generated by the bar, such that it nearly becomes co-rotating with the bar, and travels mostly radially, in $x_{1}$-type orbits, along the length of the bar. 

\begin{figure}
	\includegraphics[width=\columnwidth]{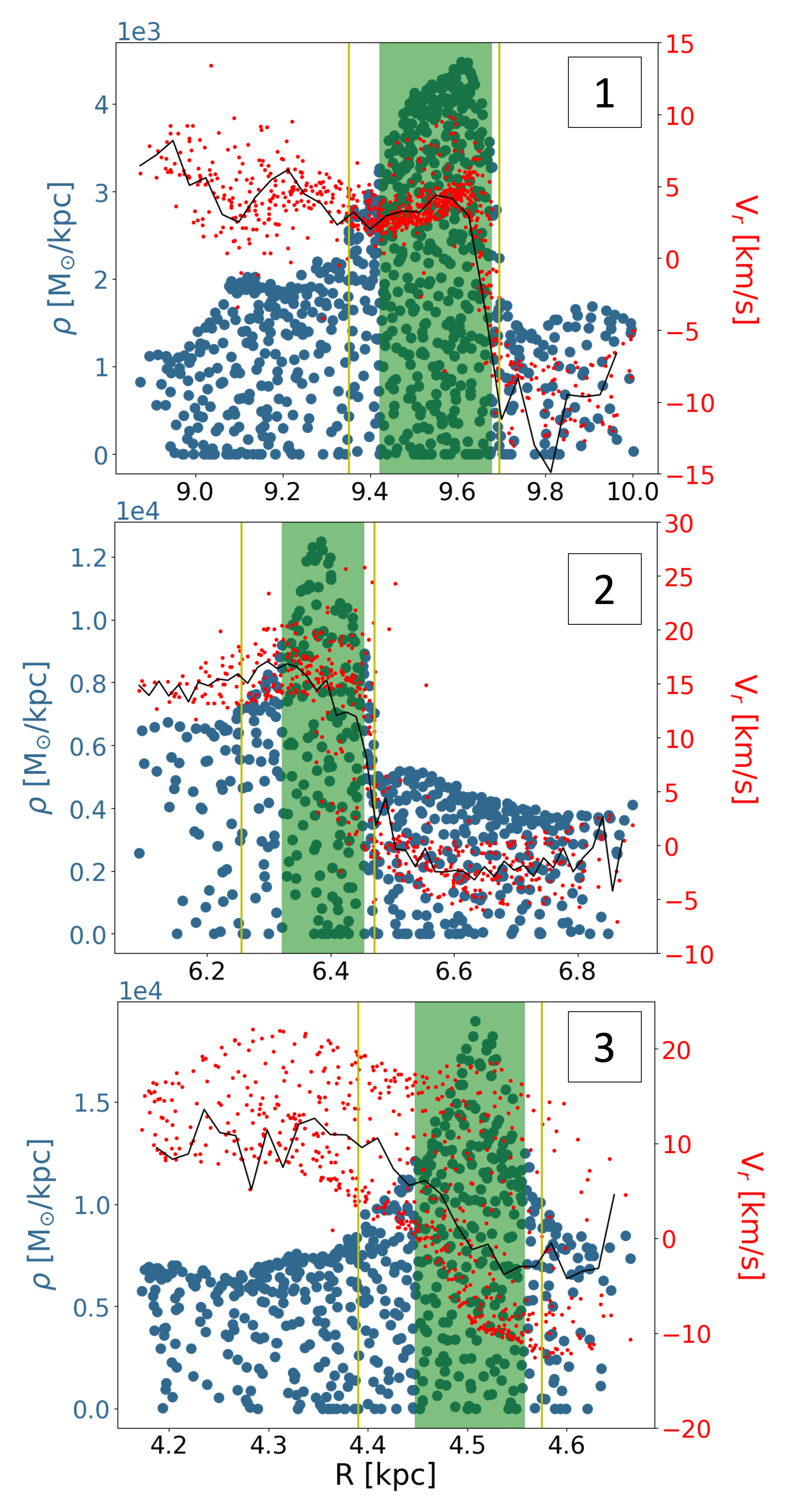}
    \caption{Spiral arm study across three different regions of  Model $4$ indicated in Fig.~\ref{fig:Velocitiesd-butterfly}, decreasing in galactocentric radius from top to bottom. Each individual image shows the variation of gaseous density along the respective lines versus the distance to the Galactic Center in dark blue, and the variation of V$_{\mathrm{rad}}$ in red. The continuous black line represents the running median of the velocity. The yellow lines are a visual indication of the width of the arm,  showing where the densities start to rise substantially above the background, whilst the green-shaded area corresponds to the full-width half maximum (FWHM) of the peak in density once substracted the background.}
    \label{fig:Vrad and arms}
\end{figure}

Lastly, we investigate the streaming motions around spiral arms from analysing the radial velocity shown in the left panel of Fig.~\ref{fig:Velocitiesd-butterfly}. We examine how the density of the gas and the radial velocity changes along a radial line that crosses three spiral arms, at various distances from the galactic centre.
The top-left panel of Fig.~\ref{fig:Vrad and arms} shows the top-view distribution of the radial velocity of  Model $4$ with black lines indicating the position of the three different spiral arms we look into. Panels enumerated from $1$ to $3$ show the variation of gas density of all cells (within a height of $\pm 300$\,pc from the galactic plane and a radius of $\pm 15$\,kpc) along those lines as a function of distance to the galactic centre in dark blue, and the variation of V$_{\mathrm{rad}}$ is shown as red dots for gas. The solid black line represents the running median of the velocities. The vertical yellow lines are positioned at the start and end points of the overdensities, corresponding to the width of the arms. The typical change in velocity across the arms can then be calculated as the difference in V$_{\mathrm{rad}}$ where the black meets the yellow lines. The green-shaded area corresponds to the full-width half maximum (FWHM) of the peak of gaseous density once substracted the background. 
The measurements for the amplitude of these streaming motions for the three studied regions, alongside the galactocentric distance of each arm segment studied, are collected in Table.~\ref{tab:ARMS}.  We can see that the typical change in radial velocity crossing a spiral arm for all three cases ranges in between $\sim 12-15$ $\mathrm{km}\ \mathrm{s}^{-1}$. However, the width and FWHM of these arms  is larger for the larger galactocentric distances. The similar velocity shift but across larger cross sections, results in effectively a lower velocity gradient across the arm, and thus resulting in weaker shocks and spiral arms that are less well defined at larger galactocentric distances. This result is only tentative, given the very low number statistics and possibly biased to our specific selection of spiral arm segments. 
However, it does suggest that the strength of the spiral arm shocks, from the streaming motions, varies as a function of position in the galaxy, and this could potentially go on to regulate the amount of dense gas (and thus regulate the conversion of atomic to molecular gas) that takes place in the arms. In order to determine what defines the strength of the spiral shock, we would need to do this across entire spiral arms.  Indeed, from Fig.\ref{fig:Vrad and arms} we can see that some spiral arms segments have strong converging motions with gas coming into the spiral arms from both sides (i.e. when we have the convergence of radially inflowing and radially outflowing gas), while other arm segments have gas entering the arm only from one direction and crossing it (i.e. either always radially moving outwards, or inwards). It would thus also be interesting to investigate if and how these dynamics may influence the ability of molecular gas to accumulate and form stars within the arms, as some extragalactic studies seem to suggest that regions with negative or positive torques (i.e. radially inflowing or outflowing gas) are not equally efficient at forming stars \citep{Meidt2013}. Doing this more comprehensive study is well beyond the scope of this paper, but is an interesting avenue of follow up work.
Nevertheless, this tentative result is consistent with observations that the Milky Way's spiral arms are better defined, with more dense and molecular gas, at lower galactocentric radii, compared to the outer arms \citep[e.g.][]{clemens1985massachusetts,bovy2015power,khanna2023measuring}.

\begin{table}
	\centering
	\caption{Spiral arm properties obtained from the density and rotational velocity of gaseous cells crossing three different arms defined in the top-left panel of Fig.~\ref{fig:Vrad and arms}: $R_{i}$ denotes the ridge position from the galactic centre of the corresponding arm, $\Delta \mathrm{V}$ is the change in velocity across the arm, $\Delta \mathrm{W}$ the width of the arm and FWHM denotes the full-width half maximum of the peak in density.}
	\label{tab:ARMS}
	\begin{tabular}{ccccc} 
		\hline
		Arm & $R_{i}$  & $\Delta \mathrm{V}$  & Width & FWHM\\
        slice     & (kpc)  & ($\mathrm{km}\ \mathrm{s}^{-1}$)  & (kpc) & (kpc)\\
		\hline
		  $1$ &     $  9.61 $&       $   \sim 14.4 $&  $ \sim 0.35 $ & 0.33 \\
            $2$ &     $  6.37 $&       $   \sim 14.0 $&  $ \sim 0.21 $ & 0.18 \\
            $3$ &     $  4.57 $&       $   \sim 12.5 $&  $ \sim 0.18 $ & 0.17 \\
		\hline
	\end{tabular}
\end{table}

\subsection{The Galactic Centre}
\label{Section:CMZ}
\begin{figure*}
	\includegraphics[width=\textwidth]{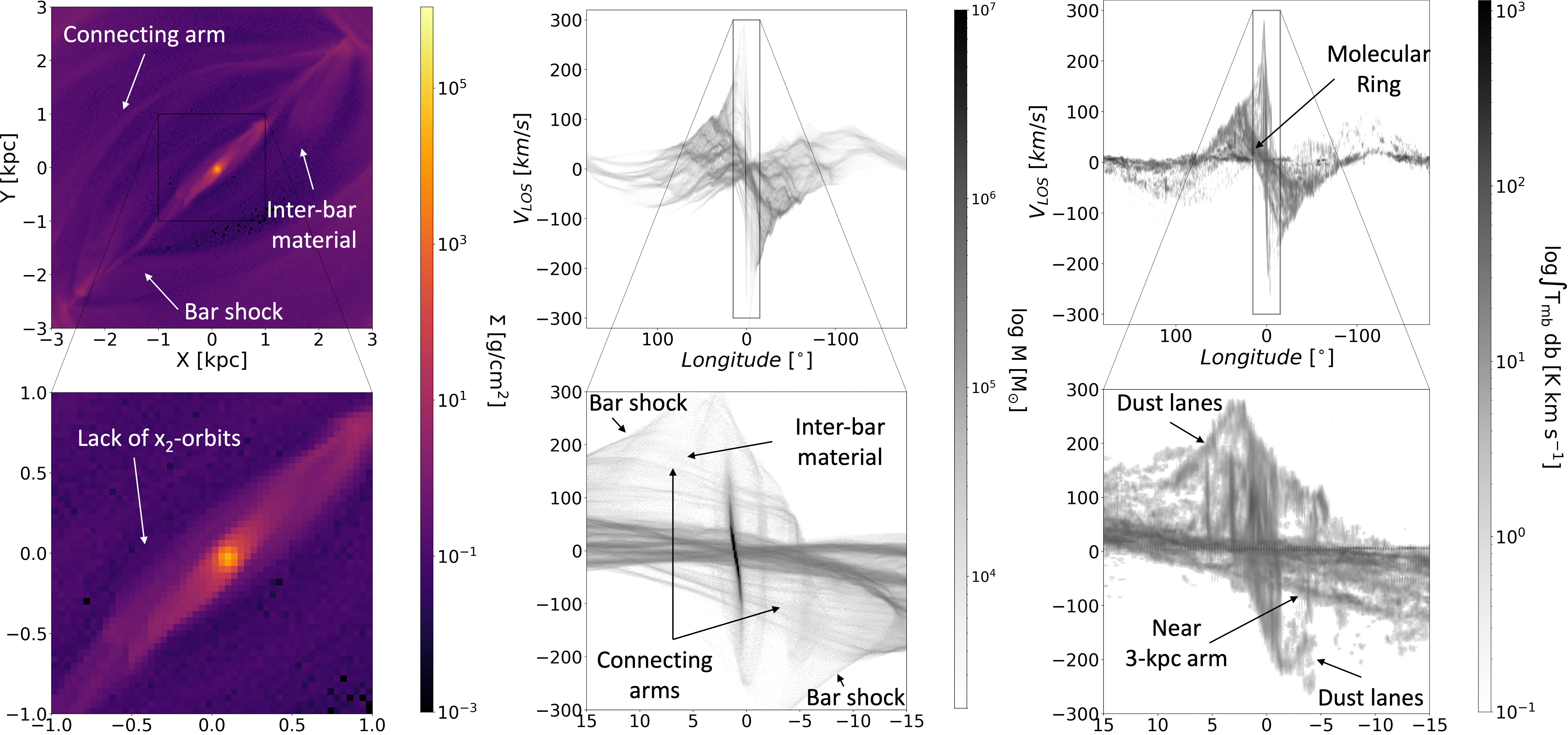}
    \caption{Top-Left: Zoom-in to the central $6 \times 6$ kpc$^{2}$ box of the top-down view of  Model $4$ at $2.6$ Gyr, where the Sun is positioned at $(0,8.2)$ kpc. Bottom-left: Zoom-in of the $2 \times 2$ kpc$^{2}$ box.  Top-middle: Longitude-velocity map of model $4$ at a viewing angle of $\phi_{obs} = 45^{\circ}$.  Bottom-middle: Zoom-in of the \textit{l-v} map of the inner $|l| = 15^{\circ}$. The proposed ``dust lane'' positions as well as the ring-like structure are marked with arrows. Top-right: $^{12}$CO \textit{l-v} map from \protect\cite{dame2001milky}. Bottom-right: Zoom-in to the observed $|l| = 15^{\circ.}$}
    \label{fig:CMZ}
\end{figure*}

We now proceed to look at the features present in the innermost parts of our best model and compare them to the $^{12}$CO observations from \cite{dame2001milky}, as well as different theoretical models. We show a zoom-in view of the central region of our  Model $4$ in Fig.~\ref{fig:CMZ}. The two left panels show a zoom-in of the central $6 \times 6$ kpc$^{2}$ box of the top-down view of the simulation on top, and the $2 \times 2$ kpc$^{2}$ box on the bottom. The Sun is positioned at $(0,-8.2)$ kpc. The middle two panels show the \textit{l-v} maps of  Model $4$ at a viewing angle of $\phi_{obs} = 45^{\circ}$. Top panel shows the entire range, whilst we zoom-in to the inner $|l| = 15^{\circ}$ on the bottom panel. The right two panels show the $^{12}$CO observations in the same range as the middle panels. 

As mentioned by \cite{binney1991understanding}, in our Milky Way, gas following $x_{1}$ orbits (parallel to the bar major axis) in the outer parts of the bar can transfer into $x_{2}$ types of orbits (perpendicular to the bar major axis) when approaching the Galactic Center. Dust lanes are dark streaks or filaments that have been observed in external barred galaxies \citep[e.g.][]{athanassoula1992existence}, and also identified in the $^{12}$CO emission in the Milky Way \citep[e.g.][]{cohen1976neutral,fux19993d, marshall2008large}. The study by \cite{pfenniger1991structure} proposed that the formation of dust lanes in barred galaxies like the Milky Way could be attributed to the intersection of $x_{1}$ and $x_{2}$ orbits. According to this model, the collision of gas clouds in these orbits creates shock waves that compress the gas and dust into a thin lane along the leading edge of the bar. More recent simulations by \cite{athanassoula2016boxy} suggest that while $x_{1}$ and $x_{2}$ orbits can play a role in shaping the gas distribution in barred galaxies, the formation of dust lanes may be more complex and involve other physical processes, such as gas inflow and turbulence.

In our case, we are able to see shock fronts that produce higher densities both in the top-view and the \textit{l-v} image of our models in the two middle panels (marked with arrows). We can also distinguish the inter-bar material with mostly non-circular motions, with gas inflowing and outflowing with respect to the Galactic Centre, influenced by the galactic bar (see also left panel of Fig.~\ref{fig:Velocitiesd-butterfly}). We can also distinguish a few ``connecting arms'' that link the two extremities of the bar, and that produce an \textit{l-v} signature similar to the observed expanding $3$-kpc arms (Far and Near components). However, we cannot distinguish $x_{2}$-type orbits in the inner parts of our  Model $4$  (see Fig.\,\ref{fig:CMZ}). These type of orbits are a result of the bar dynamics and the bulge size and mass \citep[see e.g.][]{bureau1999nature}. Some of our heavier-disc models like 6, 11 and 12, present these features in the top-view images (see Appendix~\ref{appendixA}). This suggests that we may be  underestimating the total potential felt by the gas in the inner parts of the galaxy model, perhaps simply as a consequence of not including gas self-gravity (thus the gas does not feel its own potential), or maybe due to our usage of a gas surface density profile that purely follows the stellar one, rather than one that matches the observed gas profile.   Another possible factor could be the slight underestimation of stellar mass in the centre , although we do not think this should be the main cause, given that we do explore models with larger central stellar masses, and they are worse matches to the overall structure (see also App.\,\ref{appendixD} for a more detailed discussion of the stellar distribution).

These type of orbits may appear once  we have higher resolution simulations which include gas self-gravity and feedback added to the simulation,  or with small adjustments to the gas surface density profile, in order to incorporate more material in the galactic centre  -- this will be tested in future work.

Indeed, \cite{ridley2017nuclear} studied the gas flow in the inner Galaxy using hydrodynamical isothermal simulations under an imposed potential,  and they found a central disc of gas following $x_{2}$ orbits. Following up on their work, \cite{sormani2018theoretical} generated higher-resolution (target mass of $100 \mathrm{M}_{\odot}$ compared to our $1000 \mathrm{M}_{\odot}$) non-isothermal simulations under the same potential, where they included chemistry,  and those simulations better reproduced these orbits, as well as explained the observed asymmetry of the CMZ.

\begin{figure}
	\includegraphics[width=\columnwidth]{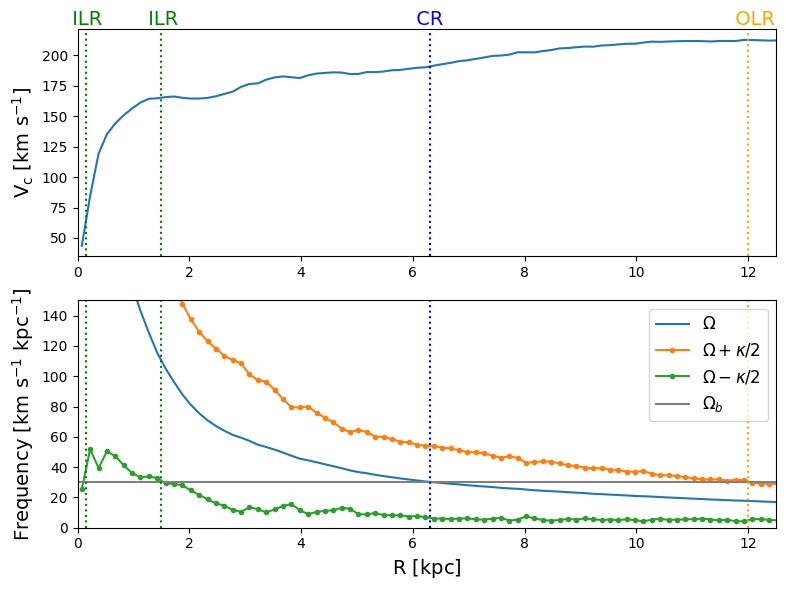}
    \caption{Top panel: the solid blue line represents the circular velocity curve as obtained from the gravitational potential in the epicyclic approximation: V$_{\mathrm{c}}(\mathrm{R}) = \sqrt{R\mathrm{d}\Phi_{o}/\mathrm{dR}}$. Bottom panel: curves for $\Omega$ in blue, $\Omega \pm \kappa/2$ in orange and green respectively. $\kappa$ represents the epicyclic frequency, $\kappa = (2\Omega / R\ )(\mathrm{d}(\Omega \mathrm{R}^{2})/\mathrm{dR}) $. The black solid line represents the value of the bar pattern speed. The ILRs occur at the intersections of the horizontal value of $\Omega_{\mathrm{p}}$ with the $\Omega - \kappa/2$ curve, CR at the intersection with $\Omega$, and OLR with the $\Omega + \kappa/2$ curve. They have been indicated with a horizontal line accross the two subplots and color-coded in accordance with the intersected curve.}
    \label{fig:Lindblad}
\end{figure}

Another option that could explain the lack of $x_{2}$-type orbits would be the position of the Lindblad Resonances. Indeed, $x_{2}$-type orbits' size and extent are known to be influenced by the location of the Inner Lindblad Resonance (ILR) in the gravitational potential, as per the epicyclic approximation \citep[e.g.][]{Contopoulus1989,athan92a}. To further investigate this issue, the top panel of Fig.~\ref{fig:Lindblad} illustrates the rotation curve derived from the gravitational potential ($\Phi_{o}$) using the epicyclic approximation: V$_{\mathrm{c}}(\mathrm{R}) = \sqrt{R\mathrm{d}\Phi_{o}/\mathrm{dR}}$. The pattern speed together with the bar-length and rotation curve suggests a slow bar in  Model $4$ with CR radius at $\sim 6.5$ kpc, which we denote in Fig.~\ref{fig:Lindblad} with a blue vertical dashed line.

 In the bottom panel, we present a resonance diagram where $\Omega$, the rotation velocity in the inertial frame, is shown in blue, and $\kappa$ is the epicyclic frequency calculated as $\kappa = (2\Omega / R\ )(\mathrm{d}(\Omega \mathrm{R}^{2})/\mathrm{dR})$. The ILRs are identified at the intersections of the horizontal black line, representing the bar's pattern speed $\Omega_{\mathrm{p}}$, with the green $\Omega - \kappa/2$ line. Co-Rotation (CR) and Outer Lindblad Resonance (OLR) are located at the intersections of $\Omega_{\mathrm{p}}$ with the blue $\Omega$ and the orange $\Omega + \kappa/2$ curves, respectively. This diagram reveals two inner ILRs, the first very close to the galactic centre at approximately $\sim 0.2$ kpc, and the second at about $\sim 1.1$ kpc. The existence of ILRs then suggests we ought to expect $x_{2}$-type orbits. However, \citet{athan92a}, in their Fig. 8, illustrate that the presence of $x_{2}$-type orbits also diminishes with an increase in bar strength. Therefore, it is plausible that in Model $4$, the bar's strength is sufficiently high to significantly reduce or even eliminate the extent of $x_{2}$-type orbits, which might explain their absence in the gas flow patterns observed. Again, in future work, with simulations with higher resolution and SN feedback, it is possible that the feedback might potentially help break out the strong bar potential, and allow these orbits to form.

The high-velocity peaks that we can see in our \textit{l-v} maps towards $l = 0^{\circ}$ can be explained as these non-circular motions that increase towards the galactic center due to the effect of the bar on the kinematics of the gas \citep[e.g.][]{fux19993d,pettitt2014morphology,pettitt2015morphology,sormani2015recognizing,li2016gas}. The strong vertical features that appear in the observations near the inner $|l| = 5^{\circ}$ are very weak to non-existent in our models. These are directly related to the high-velocity gas along the dust lane that overshoots the CMZ and is a accreted later in time (see e.g. \citealt{regan1997mass} for observations in NGC 1530 or \citealt{sormani2018theoretical} and \citealt{hatchfield2021dynamically} for theoretical modelling).  When comparing to the observations in Fig.~\ref{fig:CMZ}, we also note that our model does not reach the $\sim 300$ $\mathrm{km}\ \mathrm{s}^{-1}$ peak around $l=0^{\circ}$. As our simulation does not yet properly resolve the dynamics of the central part, it cannot reproduce these features.

The bright band of molecular gas crossing the Galactic centre in $l-v$ maps, is sometimes referred to as the "molecular ring" \citep[e.g.][]{dame2001milky}, which was first thought to be due to a resonance \citep[e.g.][]{binney1991understanding,Combes1996}. However, whether this feature actually corresponds to a physical ring is debated, and some works suggest instead that this dynamical feature is composed of the inner parts of spiral arms \citep[e.g.][]{Bissantz2003,Nakanishi2006,rodriguez-fernandex2008}. In Model 4, we observe a significant amount of material in the inner \textit{l-v} image from Fig.~\ref{fig:CMZ}, which resembles the observed structure marked in the \textit{l-v} map by \cite{dame2001milky}. However, this gas is not actually distributed in a ring-like structure in the top-down view of the galaxy, and the $l-v$ plot does show that this band is composed of multiple tracks overlaid. This supports the idea that the observed $l-v$ feature might be, in fact, arising from the inner segments of spiral arms, connected to the bar tips \citep[e.g.][]{Dobbs2012}.

Given that our models are not specifically tailored to mimic the Galactic Centre of the Milky Way, it is reassuring that our best overall model is in fact able to reproduce most of the observable features of the inner Galaxy, even though it inherits the more dynamic/uncontrolled nature of a live/dynamic stellar potential.  In addition, comparing the gas flow patterns in Model $4$ with those in the study by \cite{li2022} reveals striking similarities.  The close resemblance with our best model provides a reassuring validation of our results. The consistency between the two models underscores the robustness of our approach in simulating the gas dynamics of the Milky Way. This model will therefore serve as our base model of the Milky Way for future work, which will involve the inclusion of more sophisticated physics, including gas self-gravity, cooling and heating of the ISM, chemistry, star formation, supernova and stellar feedback, and ultimately magnetic fields, all at higher resolution.

\section{Summary and conclusions}
\label{Section: Conclusions}

We have used the moving mesh {\sc arepo} code with live stellar and dark matter potentials and isothermal conditions for the gas, in order to produce a suite of numerical simulations whose aim was to recover the large-scale structure of the Milky Way. The initial conditions have been built using the {\sc{makenewdisk}} code, whose parameters were largely based on the best fit model of \cite{pettitt2015morphology}. However, as in their study they do not obtain a bar in the galactic centre, we create a set of 15 different models varying in initial stellar mass and stellar bulge fraction.

We compare our suite of simulations with $^{12}$CO observations from \cite{dame2001milky} using \textit{l-v} maps, where we explore the galactic features that would be observed at different times, by placing the observer inside the model galaxy, and varying the Sun position using a range of angles around the galactic centre. From the \textit{l-v} diagrams, we obtain the skeletons of the main structures (Section~\ref{sec:skeletons}), and use the SMHD metric and fraction of mismatches, to obtain a best fit in time and $\phi_{obs}$ for each of the models. This analysis suggest that the set of models with lower total stellar mass are generally better fits,  with models 2, 3, 4 and 5 providing the best fits, with a slight preference for Models 2 and 4 . We then analyse the terminal velocity of our numerical simulations compared to the combined $^{12}$CO and \ion{H}{I} observations of the Milky Way. This analysis suggests that  Model $4$, is our global best fit model, at a time $\sim 2.6$\,Gyr,   when viewed through an observer placed at   $\phi_{obs}=$20-45$^{\circ}$ inclination with respect to the galactic bar.

We then proceed to characterise the details of some of the features of our best model in order to verify its ability to reproduce specific features of the Milky Way.  We first look at the galactic bar and find a bar with a half-length of $3.1 \pm 0.8$kpc, comparable to the results from recent studies using the \textit{Gaia}  EDR3 data, where they find a bar with a half-length of  $\sim 4$ kpc  \citep[see e.g.][]{queiroz2021milky}. The pattern speed for our best model is found to be $\Omega = 30.0 \pm 0.2$\,km\,s$^{-1}$\,kpc$^{-1}$, which is also within the observed range of 30-40\,km\,s$^{-1}$\,kpc$^{-1}$.
 
We also looked at the dynamics of our models by generating the top-down view of the radial and tangential velocities of the gas cells. The radial velocities show a ``butterfly'' or quadrupole pattern in the central region, typical of barred galaxies. We then analyse the amplitude of the streaming motions around the spiral arms of the model, by looking at the change in radial velocity across sections of a selection of spiral arms at different  positions. This analysis reveals that the change in velocity that gas cells experience across the arms is around $\sim 12-15$\,km\,s$^{-1}$, and the arm widths are larger for the arm segments at larger radial distance. Although from the low number statistics we cannot conclude that this galactocentric trend holds throughout the entire galaxy, this result does suggest that the effective velocity gradient in the arms (and thus the strength of the spiral arm shocks) varies across the galaxy, and that the enhancement in the densities in the arm is correlated with the strength of the velocity gradient. This effect could potentially go on to regulate the conversion of atomic to molecular amount in the arms, in different parts of the galaxy, but we do not include chemistry in these models to test this. The tentative trend of weaker spiral shocks at larger galactocentric distances that we see in our model is consistent with observations that find that the spiral arms in the Milky Way are denser and more molecular in the inner portions of the Galaxy, compared to the outer arms.

Finally, we look at the main structures of the galactic centre of our best model and compare them to the $^{12}$CO observations, as well as a number of theoretical models. We look at the inner $6 \times 6$ kpc$^{2}$ and $2 \times 2$ kpc$^{2}$ inner top-down views of our simulation, as well as the inner $|l| = 15^{\circ}$ of both observed and numerical \textit{l-v} maps. We are able to identify the bar shocks, the inter-bar material with highly non-circular orbits,  a structure in the \textit{l-v} map that resembles that of the observed so-called ``molecular ring'' that in fact correspond to the inner sections of the spiral arms starting at the tip of the bar, and lastly we can also detect a few ``connecting arms''  which produce \textit{l-v}-tracks comparable to the $3$-kpc expanding arm from observations. We do not reproduce with clarity the $x_{2}$ orbits on which gas in the CMZ in the inner few hundred parsecs of the Galaxy flows.  Even though there are two ILRs in the inner $0.1-1.1$ kpc, the resonance is too weak to support $x_2$ orbits in the non-axisymmetric bar potential, suggesting that we might lack mass in the inner disc \citep[see e.g.][]{athan92a,sormani2018theoretical}.

Given that our models are not specifically tailored to mimic the exact potential of the Milky Way, nor do they include the evolution history of the Milky Way and its past or present interactions (which could be responsible for some of the large-scale galactic features), it is remarkable that from our best overall model, which inherits the more dynamic/uncontrolled nature of a live stellar potential, is capable of reproducing many of the observable features of the Galaxy, including the global spiral arm pattern, terminal velocity and bar properties, as well as more specific features such as the amplitude of the streaming motions around spiral arms, and the dynamics of the connecting arms and bar shock material in the centre of the galaxy.   The two main caveats of our model in terms of being a good reproduction of the MW are the slight underestimation of the overall rotation curve, and the lack of $x_{2}$-type orbits in the very centre of the galaxy  (which might be solved with the inclusion of self-gravity, ISM chemistry and SN feedback in future work). This model   will thus serve as the base configuration for more sophisticated modelling of the ISM in   a MW-like galaxy - by including more physical processes such as SNe and stellar feedback, chemistry, sink particles and magnetic fields. A model that is capable of reproducing most of the Milky Way observable structure self-consistently, is therefore a powerful tool to gain insight into what might be regulating star formation in our Galaxy, giving the opportunity to increase the resolution and zoom-in to individual molecular clouds and star-forming cores in a different range of environments within a   MW-like model.

\section*{Acknowledgements}

The calculations presented here were performed using the supercomputing facilities at Cardiff University operated by Advanced Research Computing at Cardiff (ARCCA) on behalf of the Cardiff Supercomputing Facility and the HPC Wales and Supercomputing Wales (SCW) projects. ADC and EDC acknowledge the support from a Royal Society University Research Fellowship (URF/R1/191609). RSK acknowledges funding from the European Research Council via the ERC Synergy Grant ``ECOGAL'' (project ID 855130), from the German Excellence Strategy via the Heidelberg Cluster of Excellence (EXC 2181 - 390900948) ``STRUCTURES'', and from the German Ministry for Economic Affairs and Climate Action in project ``MAINN'' (funding ID 50OO2206). RSK also thanks for computing resources provided by {\em The L\"{a}nd} through bwHPC and DFG through grant INST 35/1134-1 FUGG and for data storage at SDS@hd through grant INST 35/1314-1 FUGG. RJS gratefully acknowledges an STFC Ernest Rutherford fellowship (ST/N00485X/1) and HPC from the Durham DiRAC supercomputing facility (ST/P002293/1).

\section*{Data Availability}

The maps for the top-down surface density, and \textit{l-v} projection of our best model (Model $4$) are publicly available in \url{https://dx.doi.org/10.11570/24.0088}, as well as in the FFOGG (Following the Flow of Gas in Galaxies) project website (\url{https://ffogg.github.io/})


\bibliographystyle{mnras}
\bibliography{sorted_updated_references}



\appendix

\section{Top-View Maps}
\label{appendixA}

\begin{figure*}
	\includegraphics[width=\columnwidth]{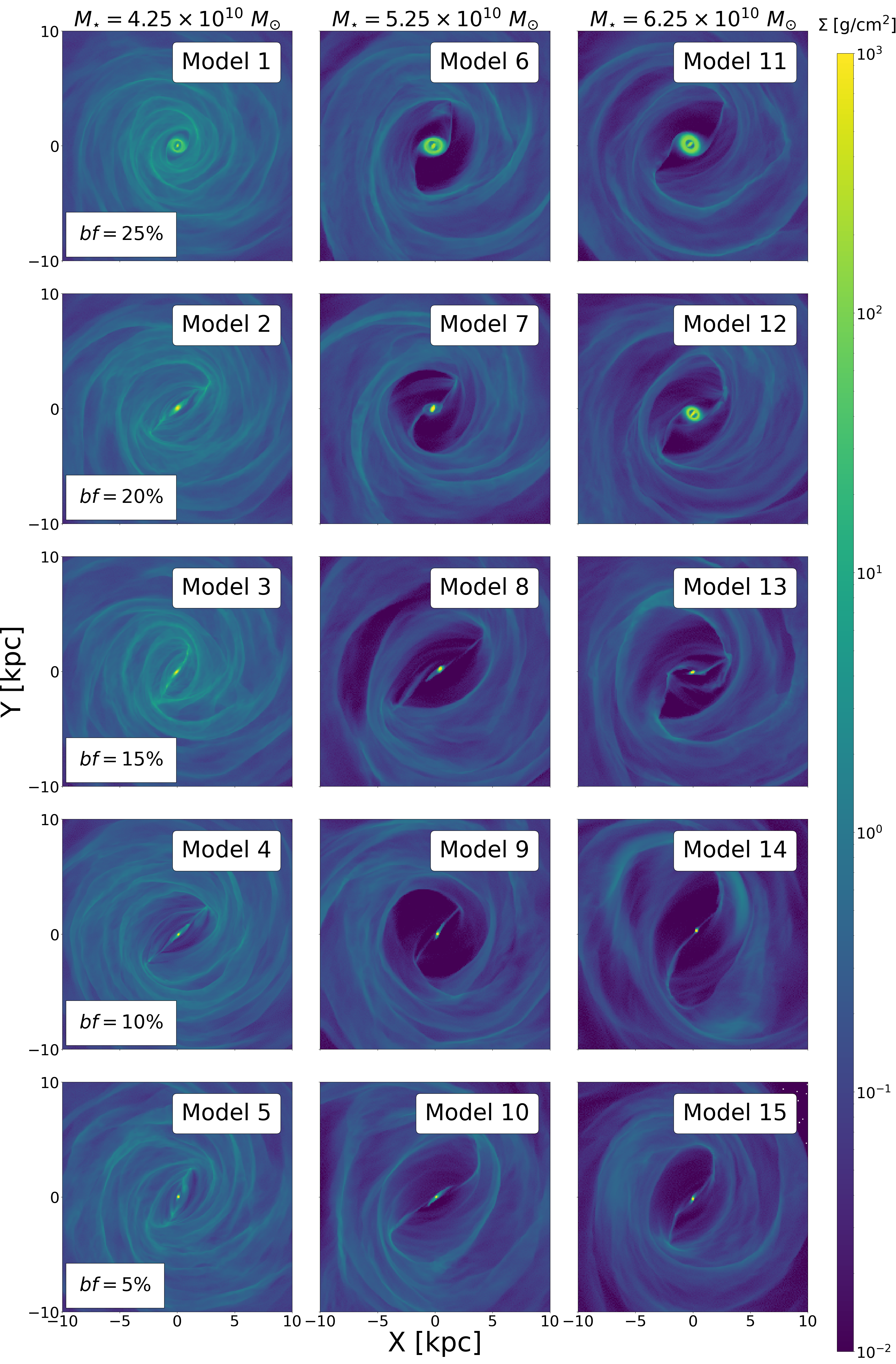}
    \hfill
 	\includegraphics[width=\columnwidth]{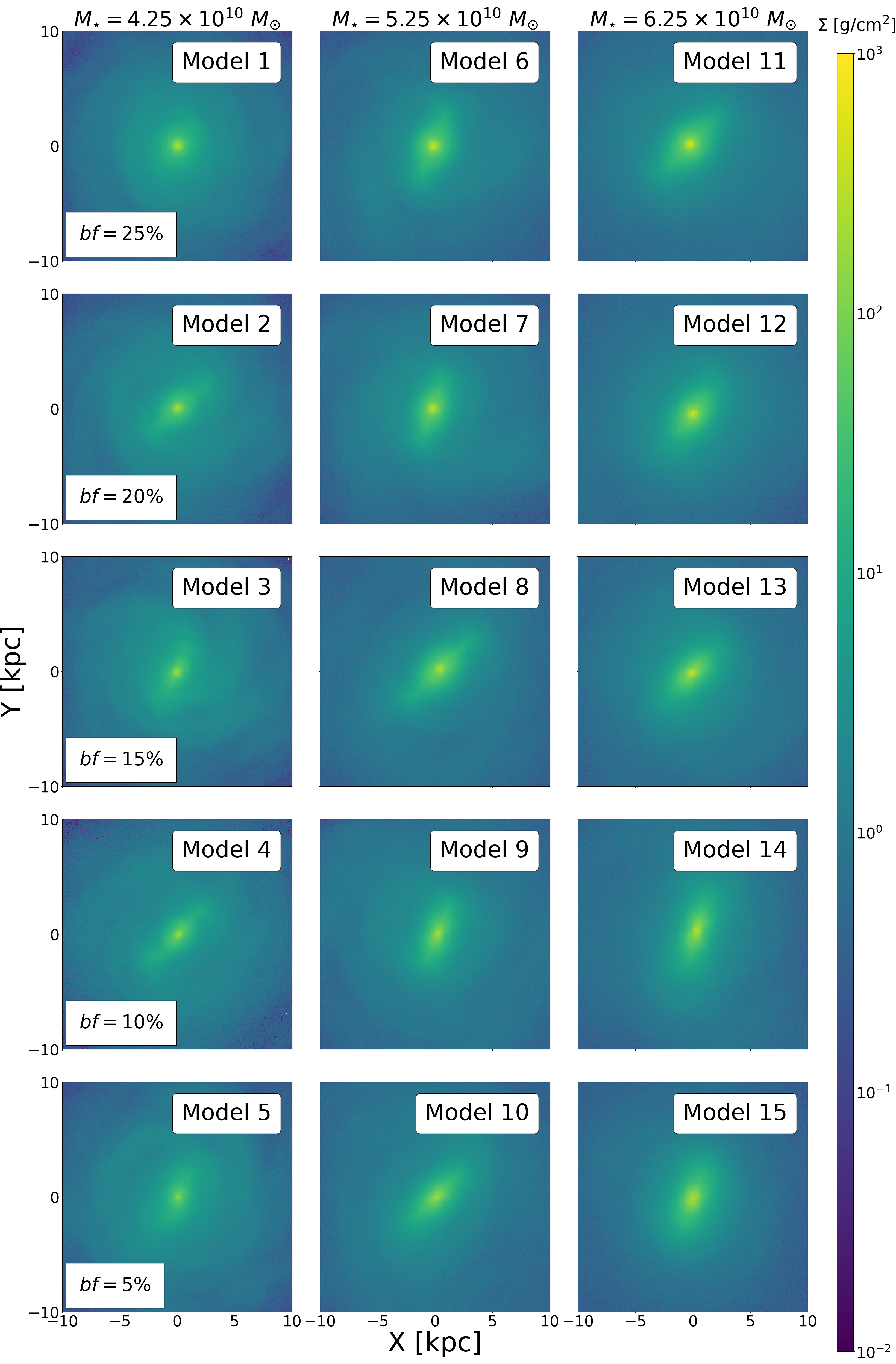}
    \caption{Top-view plots of the inner $10 \times 10$ kpc$^{2}$ box for the gas distribution (left) and stars (right) of all $15$ models that make up our sample, at the optimal time for each model and with the Sun positioned at (0,-8.2) kpc. The initial stellar mass increases from left to right and the bulge fraction decreases from top to bottom. The colorbar represents the surface density in g/cm$^{2}$.}
    \label{fig:top-view gas}
    \label{fig:top-view stars}
\end{figure*}

Figure.~\ref{fig:top-view gas} shows the top-down view plots of gas (left) and stars (right) for the inner $10 \times 10$ kpc$^{2}$ box of our different $15$ models at their optimal time, as selected from the SMHD metric (see Section \ref{sec:SMHD_analysis}). The initial stellar mass increases from left to right and the bulge fraction decreases from top to bottom. The colour scale is the same for all panels and represents the surface density in g\,cm$^{-2}$. All models, except Model 1, produce a bar in the galactic centre. Bars can be seen in both stellar and gas distributions. The observer's viewing angle differs between models and is explained in more detailed in Section.~\ref{sec:lv-plots}. 

\section{Fraction of mismatches}
\label{appendixB}

The fraction of mismatches that we introduce in Sect.\,\ref{Section: SMHD} is obtained using Eqs.~\ref{eq:mismatch} and \ref{eq:global}. The goal of this metric is to quantify the number of structures in the \textit{l-v} maps that our simulations create in excess (or in deficit) when compared to observations. For each pixel in our simulations, a distance to the closest point in the observations is calculated, and vice versa. However, if, for example, the \textit{l-v} map of a given model contains most structures, by construction, it will mean that when looking for the closest point from the observations to that model, the closest points will always have small distances (and therefore giving a low SHMD metric value), while the model is not actually a good representation of the observed structures. Therefore, values for our metric SMHD can be underestimated and favour models that do not necessarily reproduce the structures of the observations. 

\begin{figure}
	\includegraphics[width=\columnwidth]{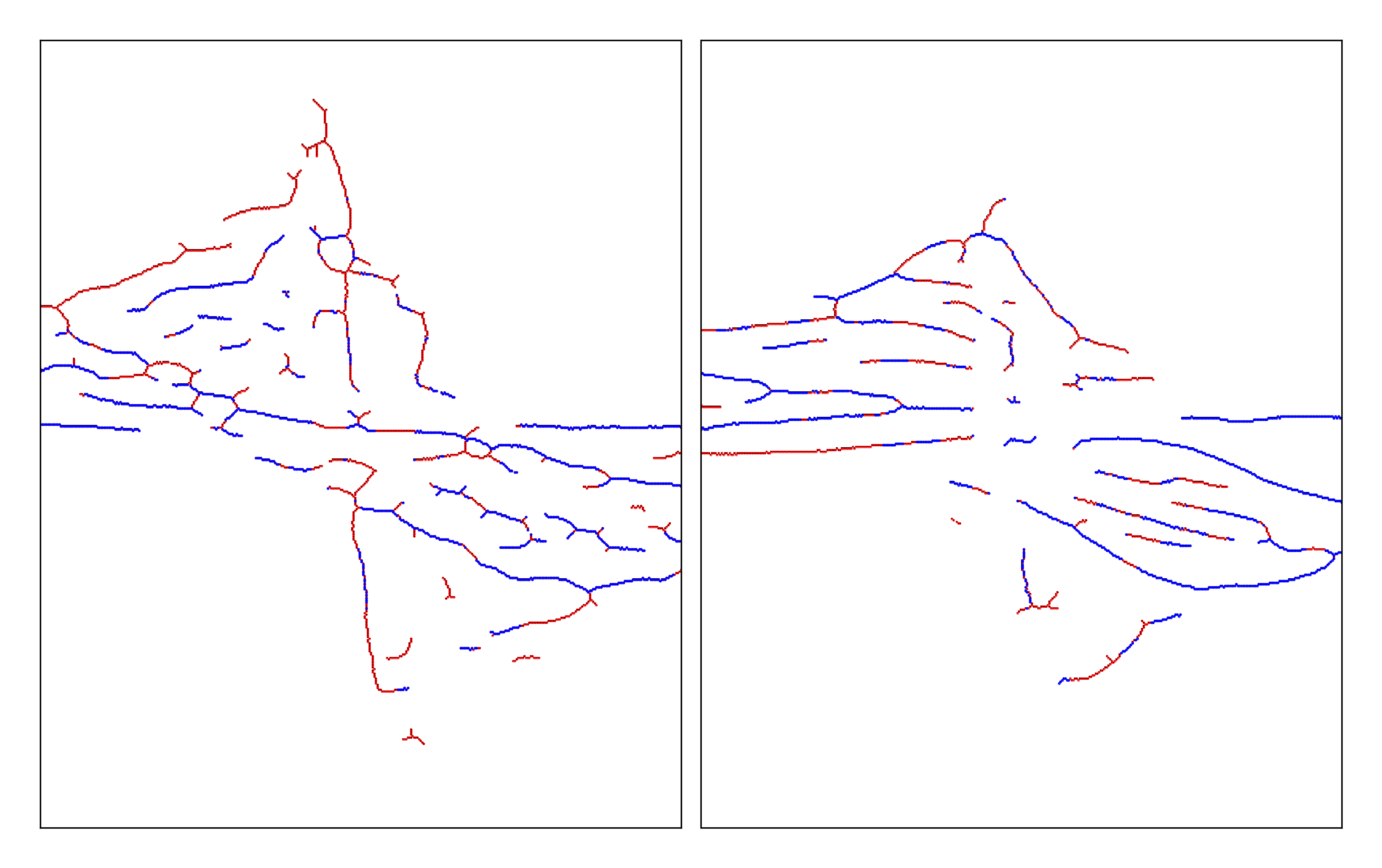}
    \caption{Skeletons extracted from the \textit{l-v} maps of the $^{12}$CO observations by \citet{dame2001milky} (left) and our Model 4 at a time $\sim 2.6$ Gyr and angle $\phi_{obs} = 45^{\circ}$ (right). Both images show the  inner $|l|=20^{\circ}$ of the Galaxy. Blue pixels are uniquely associated with each other, whilst red-coloured ones display those defined as mismatches.}
    \label{fig:skeleton_mismatches}
\end{figure}

This effect is illustrated in Fig.~\ref{fig:skeleton_mismatches}. 
Both panels show a zoom-in of the inner $\sim 20^{\circ}$ region of the Galaxy. The figure shows the skeletons of the \textit{l-v} maps of the $^{12}$CO observations by \cite{dame1986largest} (left) and our model 5 at a time $\sim 2.3$ Gyr and angle $\phi_{obs} = 150^{\circ}$ (right). Pixels are colour-coded depending on whether they are unique pairs in blue (i.e. pairs which are common on both directions: sims-to-obs and obs-to-sims), and pixels classified as mismatches in red. We can see that a large number of skeletons in red in the model do not correspond to any structures in the observations. These contribute to lowering the SMHD metric, but the model does not represent a good match of the observations. This is the reason why we introduce the extra metric of the fraction of mismatches, where only the pairs of pixels which are common on both directions (i.e. that appear in both sims-to-obs and obs-to-sims), are considered to be a good match of the underlying structure. The lower the fraction of mismatches, the better the match. We use this metric in conjunction with the SMHD metric (i.e. the best overall fit of the observed structure are the models with low values on both metrics). 

\section{Outline of the \textit{l-v} maps}
\label{appendixC}

\begin{figure}
	\includegraphics[width=\columnwidth]{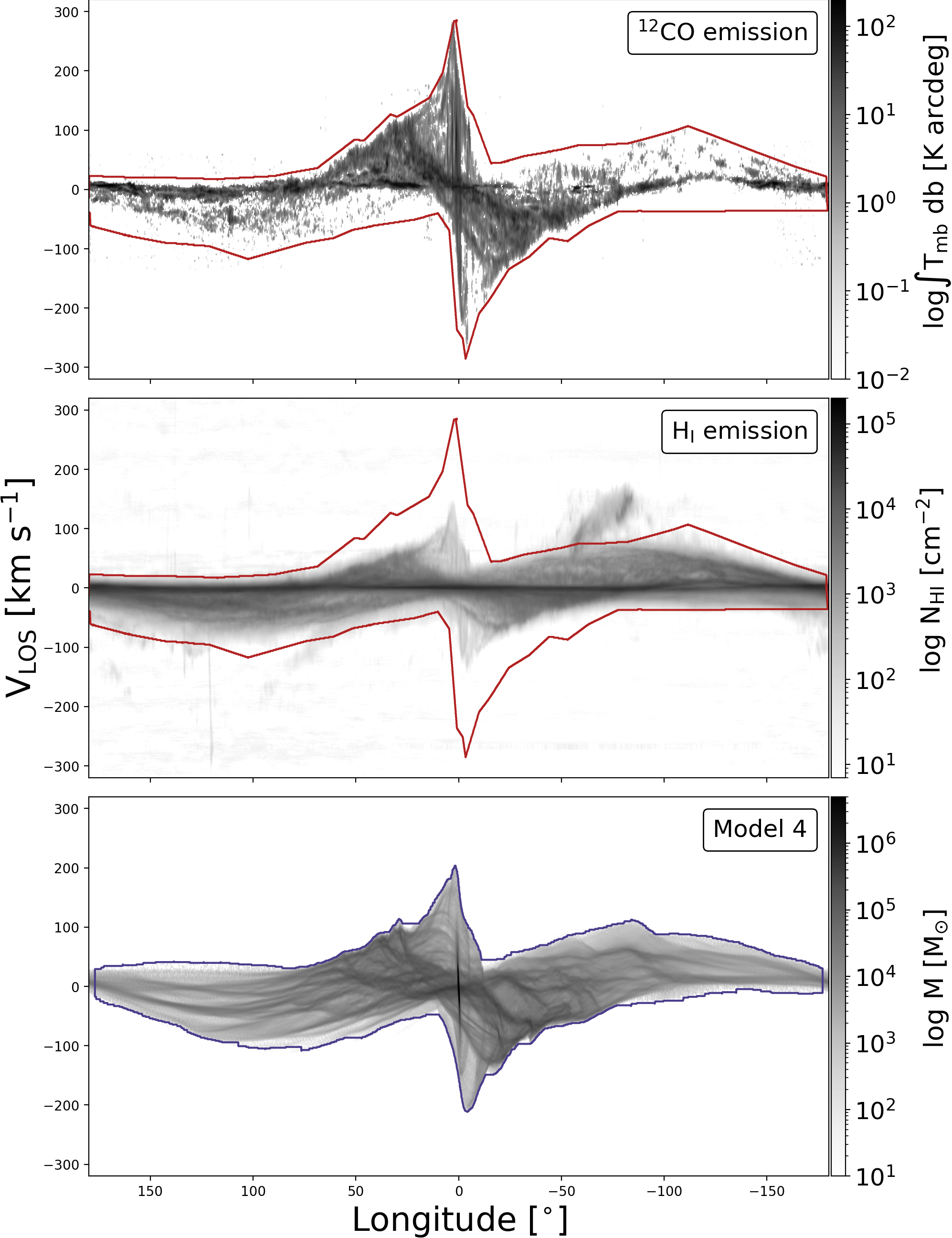}
    \caption{Top: \textit{l-v} map of the $^{12}$CO emission from the work by \protect\cite{dame2001milky}. Middle: \textit{l-v} map of the \ion{H}{I} emission extracted from the HI-4PI Survey \protect\citep{bekhti2016hi4pi}. Bottom: \textit{l-v} map from our  Model $4$ at a time of $\sim 2.6$ Gyr. All images have been superimposed with their respective outlines (red continuous line for observations, blue for simulation) used for the analysis of the terminal velocity (see Section~\ref{sec: terminal velocity} for methodology and Section.~\ref{results:term-vel} for results and discussion).}
    \label{fig:outline}
\end{figure}

Figure.~\ref{fig:outline} shows the \textit{l-v} maps of the $^{12}$CO and \ion{H}{I} emission (top and middle panels respectively) and our  Model $4$ at a time of $\sim 2.6$\,Gyr and viewing angle $\phi_{\mathrm{obs}}=45^{\circ}$ (bottom panel). The selected outline for the terminal velocity has been superimposed on each of the images: red solid line corresponds to the combined outline from both observations, and blue contour shows the silhouette of the \textit{l-v} map of our model (see Section\ref{sec: terminal velocity} for more details on the methodology). Colours have been chosen to match those of Fig.~\ref{fig:outer_edge} for consistency. Note that in the middle panel of Fig.~\ref{fig:outline} there is some emission not included within the observational contour at longitude $\sim -80^{\circ}$ and velocity $\sim 100$\,$\mathrm{km}\ \mathrm{s}^{-1}$. This is \ion{H}{I} emission from the Magellanic Clouds and therefore it is excluded for the study of the terminal velocities of the Milky Way.

\section{Stellar distribution}
\label{appendixD}

  This Section presents a brief overview of the main properties of the stellar distribution of our best model. A more in-depth analysis of these and their direct comparison to observational studies will be subject of a follow up study.

\begin{figure}
	\includegraphics[width=\columnwidth]{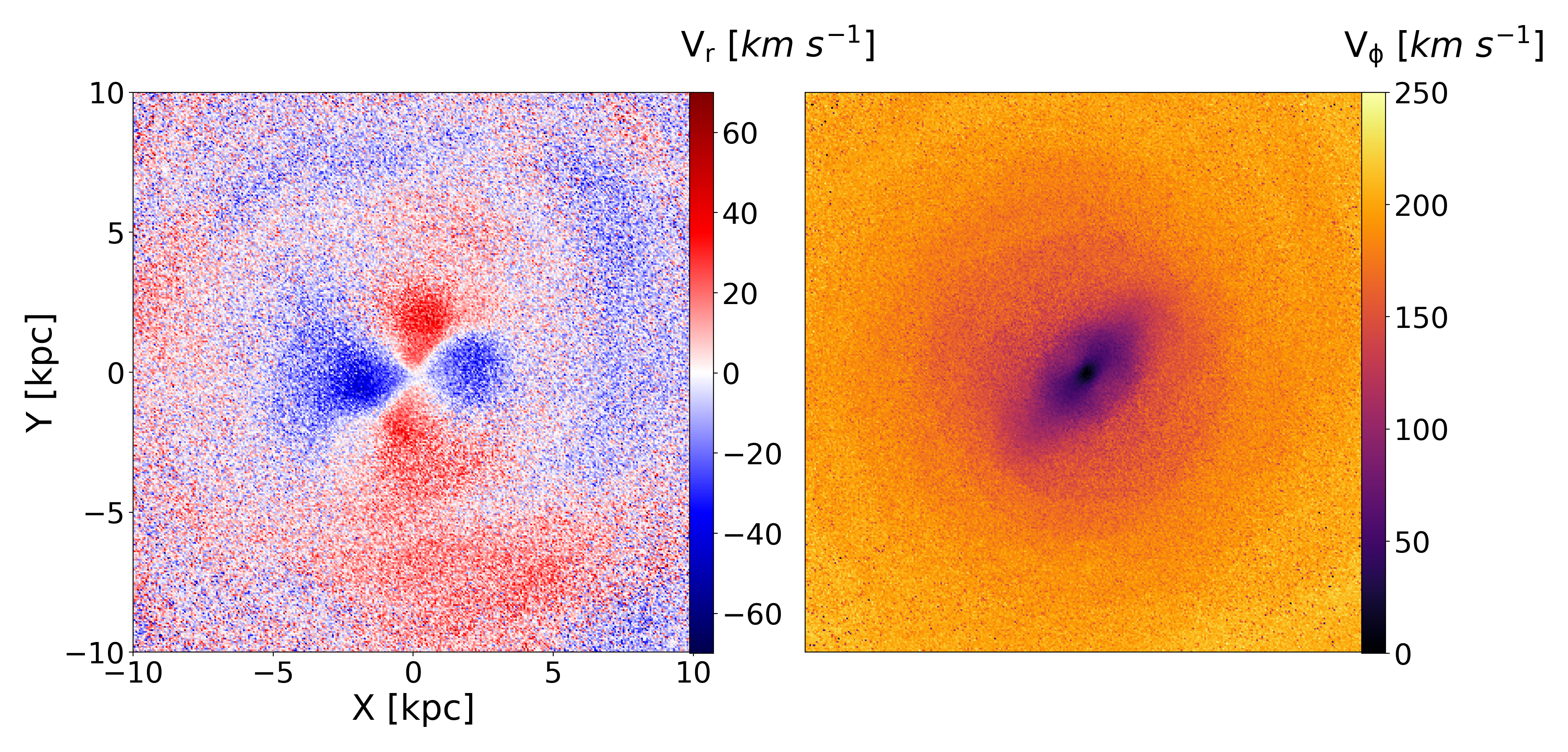}
    \caption{ Top-down view of the $20 \times 20$\,kpc$^{2}$ stellar distribution of Model 4 (same as Fig.~\ref{fig:Velocitiesd-butterfly} but for the stars). The left panel is colour-coded with the radial velocity, V$_\mathrm{rad}$, of stars, while the right panel shows the tangential component, V$_{\phi}$.}
    \label{fig:Velocity-butterfly-stars}
\end{figure}

\begin{figure}
	\includegraphics[width=\columnwidth]{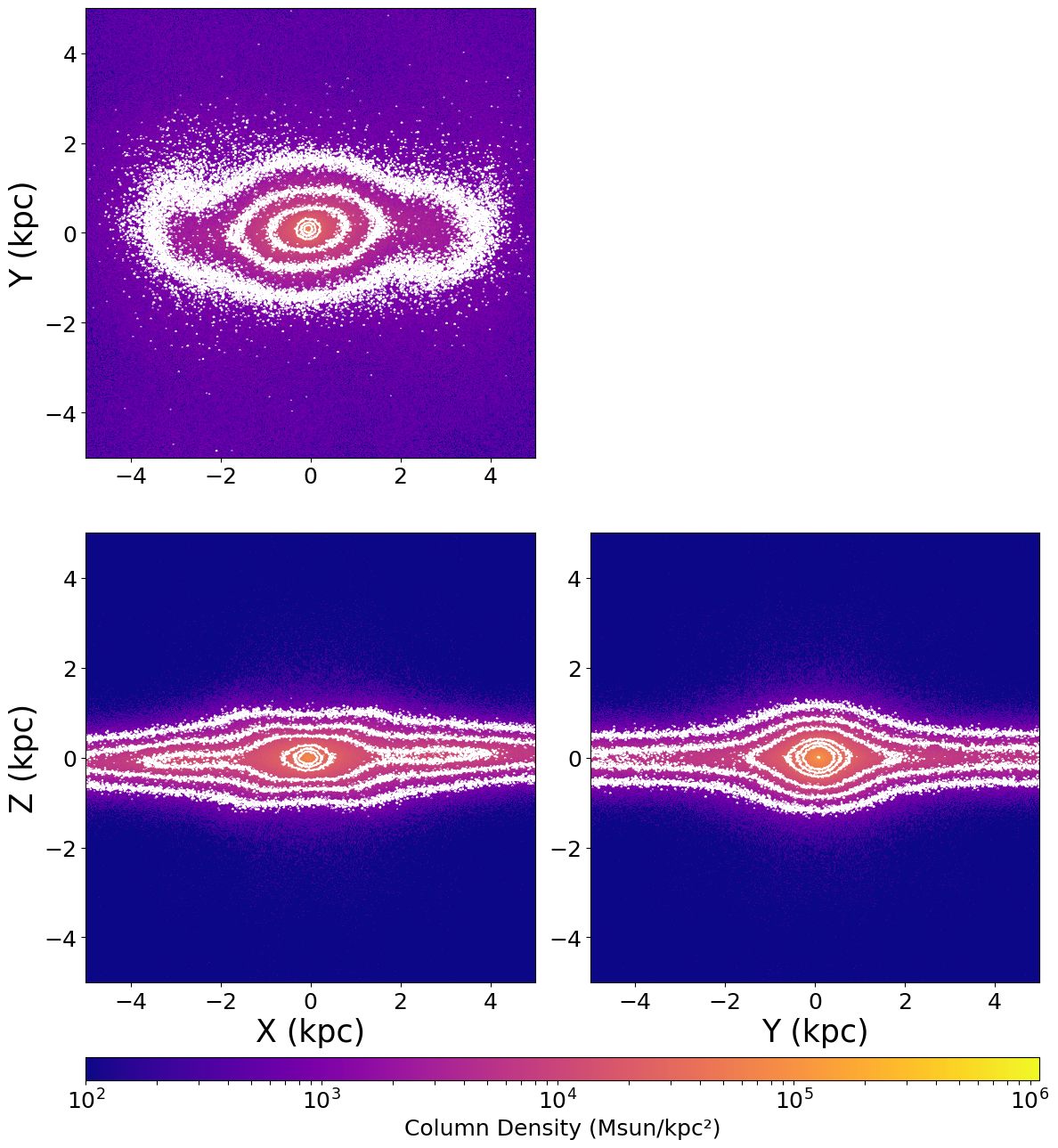}
    \caption{ Stellar distribution within a 10 × 10 x 10 kpc$^{3}$ region of  Model $4$ at approximately $\sim 2.6$ Gyr, showcased in $xy$ (top), $xz$ (bottom-left), and $yz$ (bottom-right) panels. White lines represent iso-density contours at 10 $\%$, 20 $\%$, 50 $\%$, 70 $\%$ and 90 $\%$ of the peak. This shows that the central area exhibits a longer fainter stellar bar (extending to $\sim3$\,kpc radius), and a more compact ellipsoidal towards the central region, with a flattened profile in the $z$ direction. }
    \label{fig:bulgy}
\end{figure}

Figure.~\ref{fig:Velocity-butterfly-stars} shows the stellar velocity fields in terms of radial and tangential velocity components. We can see that these mimic the pattern seen for the gas (from Fig.~\ref{fig:Velocitiesd-butterfly}), albeit with less sharp transitions. We can see that the radial velocity field exhibits a quadrupole pattern, comparable to results found in observations of the Milky Way with APOGEE-Gaia \citep{leung2022} or \textit{Gaia}-DR3 \citep{drimmel2023}, albeit perhaps the pattern in our model has a slightly smaller extent.

 Figure.~\ref{fig:bulgy} shows the stellar distribution within a $10 \times 10 \times 10$\,kpc$^{3}$ region of  Model $4$ at approximately $\sim 2.6$ Gyr, showcased in $xy$ (top), $xz$ (bottom-left), and $yz$ (bottom-right) panels. White lines represent iso-density contours, aiding in visualising the distribution. We can see that this model develops a long bar and a more ``boxy-peanut'' shape of the inner stellar distribution ($R<1-2$\,kpc), in line with some observational studies of the MW stellar bar \citep[e.g.][]{benjamin2005,wegg2015structure}. 

 We also estimate the total stellar mass enclosed within a 5\,kpc distance from the galactic centre, and up to 1.5\,kpc off the midplane (equivalent to the observations), and obtain a total enclosed stellar mass of $2.65\times10^{10}$\,M$_{\odot}$ for the initial time of Model 4, and $1.98\times10^{10}$\,M$_{\odot}$ for the best time. These values are notably close to the observed MW estimates provided by \cite{Valenti2016} of $2.0\pm0.3\times10^{10}$\,M$_{\odot}$ , but slightly lower than those estimated by \cite{portail2017dynamical} of $3.17\pm0.24\times10^{10}$\,M$_{\odot}$.

\section{\textit{L-V} tracks on observations}
\label{appendixE}

\begin{figure*}
	\includegraphics[width=\textwidth]{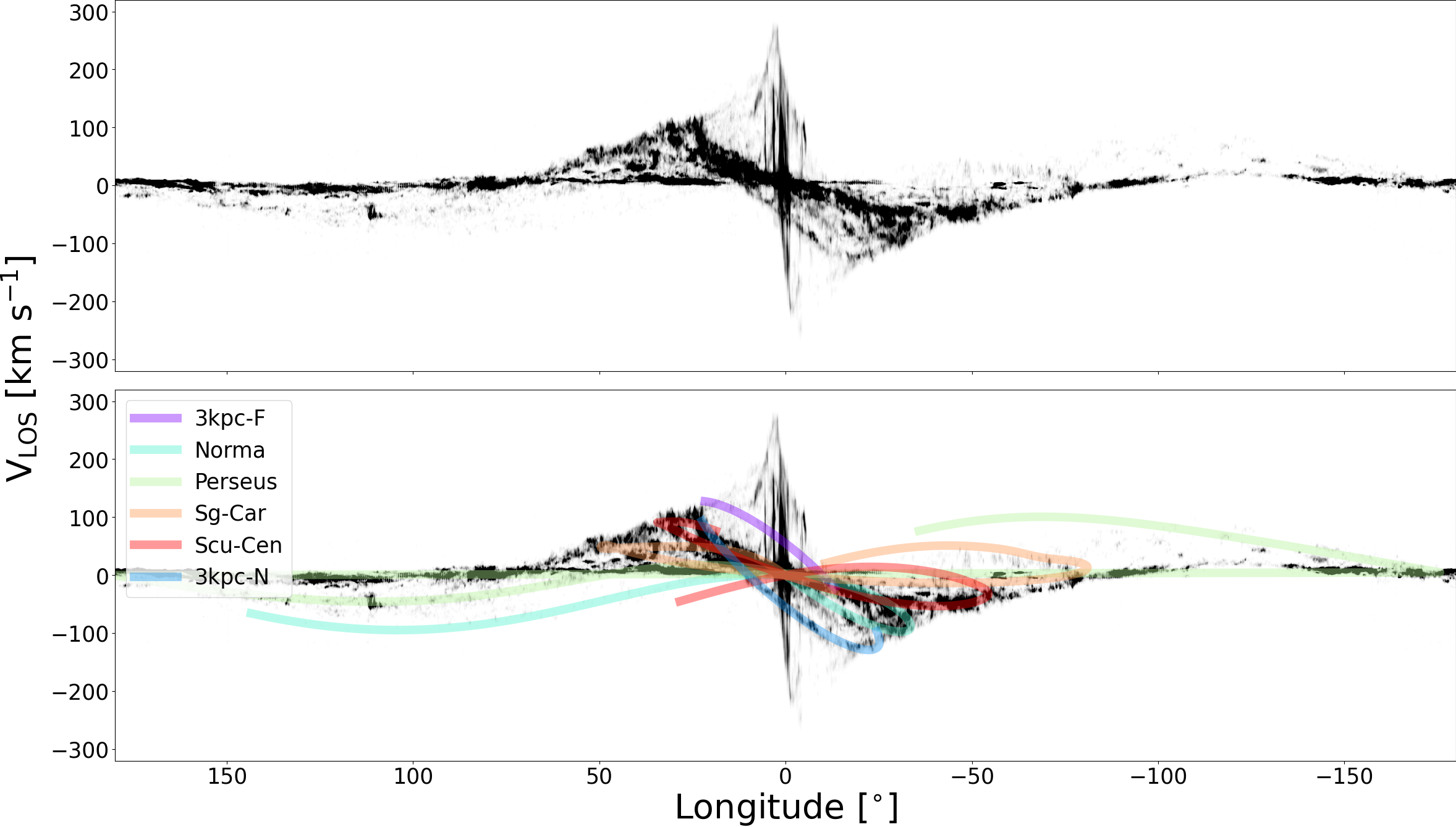}
    \caption{Same as Fig.\ref{fig:Spiral Pattern -LV} for the observations. Top: $l-v$ map of the $^{12}$CO observations by \citet{dame2001milky}. Bottom: Same image, with the spiral arms tracks from \citet{taylor1993pulsar} superimposed. These tracks are displayed with a width of 10\,$\mathrm{km}\ \mathrm{s}^{-1}$ and the different colours represent different spiral arms: the 3\,kpc-arm is represented in blue for the Near and purple for the Far counterparts; and light blue, light green, yellow and orange represent the Norma-Outer, Perseus, Sagittarius-Carina and Scutum-Centaurus arms respectively.}
    \label{fig:Spiral Pattern -LV-CO}
\end{figure*}

Figure.~\ref{fig:Spiral Pattern -LV-CO}  presents the \textit{l-v} map based on the $^{12}$CO observations by \citet{dame2001milky} arranged similarly to Fig.\ref{fig:Spiral Pattern -LV} for visual comparison. Additionally, the bottom panel of Figure.~\ref{fig:Spiral Pattern -LV-CO} incorporates the same spiral arm tracks obtained from 
\citet{taylor1993pulsar}.


\bsp	
\label{lastpage}
\end{document}